\documentclass[preprintnumbers,showpacs,amsmath,amssymb,floatfix,9pt,prd,onecolumn,
superscriptaddress,nofootinbib]{revtex4}
\usepackage{latexsym}
\usepackage{epsfig}
\usepackage{amssymb}
\begin{document}

\title{\textbf{Recovery of hadronic cross section inside the mini black holes at LHC}}

\author{Tooraj Ghaffary}
\email{ghaffary@iaushiraz.ac.ir} \affiliation{Department of Science,~Shiraz Branch,~Islamic Azad University, Shiraz, Iran}

\author{Alireza Sepehri}
\email{alireza.sepehri3@gmail.com} \affiliation{Faculty of Physics, Shahid Bahonar University, ~ P.O. Box 76175, Kerman,Iran\\
Research Institute for Astronomy and Astrophysics of Maragha (RIAAM),~ P.O. Box 55134-441, Maragha, Iran}

\author{Sayeedul Islam}
\email{sayeedul.jumath@gmail.com} \affiliation{Department of Mathematics,Jadavpur University, ~ Kolkata-700 032,West Bengal,India}

\author{Farook Rahaman}
\email{rahaman@associates.iucaa.in} \affiliation{Department of Mathematics,Jadavpur University, ~ Kolkata-700 032,West Bengal,India}

\begin{abstract}

In curved space time inside the shell of TeV black holes, many gluons and quarks produce due to Unruh effect, interact with each other and create Higgs boson. We study the Unruh effect and show that the inside state of the black hole for gluons and quarks can be represented by a maximally entangled two-mode squeezed state of inside and matter Hilbert spaces of black hole.  We consider different channels for Higgs boson production inside the shell and outside the shell of mini black holes at LHC and obtain its cross section in each channel. Comparing the Higgs boson production outside and inside the shell ,we observe that at lower mass $“M_{BH}<4TeV”$ and at higher mass $“M_{BH}>9TeV”$, the  Higgs boson will not be produced outside the shell of black hole,However this particle can be produced inside the shell for $“2.5TeV<M_{BH}<12.5TeV”$.Also comparing Higgs production via quark or gluon interaction we find that the outside the shell of black hole is able to produce  a quark for $“3TeV< M_{BH}<10TeV”$; and eventually for $“M_{BH}<3TeV and M_{BH}>12.5TeV”$ the black hole can only emit massless  gluons.However inside the shell of black hole is able to produce a quark for $“2.5TeV< M_{BH}<14TeV”$.Comparing these Higgs boson cross sections with Higgs boson cross sections in Perturbative Quantum Chromo Dynamics $(PQCD)$, we find that the micro black hole can be a source for Higgs production at LHC. Finally we calculate the effects of Higgs boson radiation due to inside the shell and outside the shell of mini black holes on hadronic cross section at LHC. We observe that as the order of perturbation theory increases, this effect becomes systematically more effective, because at higher orders there exists more channels for Higgs production and their decay of Higgs into massive quark-anti quarks in our calculation. At smaller mass, $(M_{BH}<2TeV)$,the NNLO contribution is large while the cross sections at NLO and at LO are rising at $M_{BH}<3TeV and M_{BH}>4TeV$ respectively and exhibit a turn-over at moderate values of  black hole mass. The peak moves from about 5TeV (LO) to 2.5TeV (NNLO). These results are different from references due to effects of Higgs boson production and decay inside the event Horizon.It$'$s concluded that the processes of hadronization inside the event horizon of mini black holes is affected the hadronic cross section outside the event horizon and can be observed at LHC.

\end{abstract}

\maketitle


\section{introduction}

In a previous paper, the effect of Higgs boson radiation from  TeV  black holes on hadronic cross section at LHC has been investigated[1]. However ,the process of hadronization  inside the event horizon is ignored.  Unfortunately the information transformation from collapsing matter to outside of event horizon of mini black holes at LHC is not complete and information is lost[2]. It is concluded that all of hadrons produced inside the black hole can not come out of it and consequently their cross sections do not observe at LHC. In this research we investigate this subject.\\
\indent Recovery information inside the black hole at the LHC may change the way and we can search for new particles. Decays of BHs, tagged with prompt leptons or photons, offer low-background environment for searcheing of new particles. For example, a 130 GeV SM-like Higgs boson may be observed with the significance of five standard deviations in one hour, day, month, or year of the LHC operation for the values of the fundamental Planck scale of 1, 2, 3, and 4 TeV, respectively[3].These Higgs bosons may be produced inside or outside the black hole .The effect of Higgs boson inside the black hole changes the observed cross section at LHC.\\
\indent As the black hole masses at the LHC are relatively small (3-7 TeV)[4,5] and also the temperatures of the black holes are very high (1 TeV), the black holes can be a source for top quark production[4] and Higgs production[5] via Hawking radiation. In fact there can be an enormous amount of heavy (super symmetry and Higgs) particles production from black holes [6], which is much more than expected from normal PQCD processes [7]. This comes about from two competing effects as the Planck scale increases:\\
1) top quark and Higgs boson productions from black holes increases because the temperature of the black holes increase as the Planck scale increases for fixed black hole masses[4,5].\\
2) The cross section for black hole production decreases[8,9].\\
\indent The outline of the paper is as following.In section II we study the Unruh states for gluons and quarks inside the mini black holes at LHC. We discuss different channels for Higgs boson production inside micro black holes at LHC in section III.In section IV we compare the cross section for Higgs production from inside the black holes and from PQCD models. And    finally we calculate the effect of the process of hadronization inside the black hole on observed cross section in section V. 


\section{The Unruh states for gluons and quarks inside the mini black holes at LHC}

In this section we extend the results of the derivation of Unruh state for Scalar[10]and Dirac fields[11] inside the black hole to the quarks and gluons.We show that the ground state for QCD matter is a maximally entangled two-mode squeezed state inside the matter Hilbert spaces of black holes.
 Let us consider the gluon quantization, using the transverse free field operator[12]

\begin{equation}
A^a_\mu=\int\frac{d^{3k}}{\left( 2\pi\right) ^{3}\sqrt{2\omega\left( k\right) }}\sum_{\lambda=1}^{3}\varepsilon^\lambda_\mu\left[ a^a_\lambda\left( k\right) e^{-ikx}+a_{\lambda}^{a\dagger}\left( k\right) e^{ikx}\right] 
\end{equation}
 
where $\varepsilon^\lambda_\mu$ are the polarization vectors satisfying the transversality condition
\begin{equation}
k^{\mu}\varepsilon^{\lambda}_\mu=0
\end{equation}
which follows directly from the Lorentz condition.The gluon field satisfies the wave equation[13]:

\begin{equation}
\left( -g\right) ^\frac{1}{2}\frac{\partial}{\partial{x}^{\mu}}\left[ g^{\mu\nu}\left( -g\right) ^\frac{1}{2}\frac{\partial}{\partial{x}^\nu}\right] A^a_{\rho,s}=0
\end{equation}

where the upper index $( a=1,...8)$ is related to eight color of gluons,the lower index $ s” (s=1,s_z=+1,-1)$ denotes the spin of gluon, $ρ (ρ=1,...d)$ is the vector index ,d is the number of dimensions and $g^{\mu\nu}$ is the metric tensor. To obtain the Unruh state for gluons,we solve the equation(3) by using collapsing shell metric. The collapsing shell metric in two-dimension is given by [10]

\begin{equation}
ds^2=
\begin{cases}
-d\tau^2+dr^2~~~~~~~~~~~~~~~~~~~~~~~~~~~~~~~~~~~~~~~~~~~~~~~~ r<R\left(\tau \right) 
\\-\left(1-\frac{2M_{BH}}{r} \right) dt^2+\frac{dr^2}{\left( 1-\frac{2M_{BH}}{r}\right) }~~~~~~~~~~~~~~~~~~~~~~~ r>R\left(\tau \right) 

\end{cases}
\end{equation}
where the shell radius R defined by:

\begin{equation}
R(\tau)=
\begin{cases}
R_0~~~~~~~~~~~~~~~~~~~~~~~~~\tau<0
\\R_0-{\nu\tau}~~~~~~~~~~~~~~~~~~~r<R\left(\tau \right) 
\end{cases}
\end{equation}

The advanced and retarded null coordinates are defined by
\begin{eqnarray}
V=\tau+r-R_0, U=\tau+r-R_0 \nonumber\\
v^*=t+r-R_0^*,  u^*=t-r^*+R_0^*
\end{eqnarray}

with $R_0^*=R_0+2M_{BH}ln(\frac{R_0}{2M_{BH}}-1)$. The null coordinates are chosen such that the shell begins to collapse at $U=V=u^*=v^*=0$. In these coordinates, the metric is given by[10,11]
\begin{equation}
ds^2=
\begin{cases}
-dUdV~~~~~~~~~~~~~~~~~~~~~~~~~~~inside~the~shell ~~~~~~~~~~~~~~~~~~~~~~r<R(\tau)
\\ -\left(1-\frac{2M_{BH}}{r} \right)du^*dv^*~~~~~~~~~outside~the~shell ~~~~~~~~~~~~~~~~~~~~~r>R(\tau) 
\end{cases}
\end{equation}

after some mathematical manipulations,we obtain near the sell surface[10]

\begin{eqnarray}
v^*=-4M_{BH}ln\left(1-\frac{vV}{\left(1-v \right)\left( R_0-2M_{BH}\right)   } \right)\nonumber\\
u^*=-4M_{BH}ln\left(1-\frac{vV}{\left(1+v \right)\left( R_0-2M_{BH}\right)   }\right)
\end{eqnarray}

where $M_{BH}$ is the black hole mass.The Penrose diagram of a collapsing star [14] is shown in Fig. 1.

\begin{figure}
\includegraphics[scale=1.3]{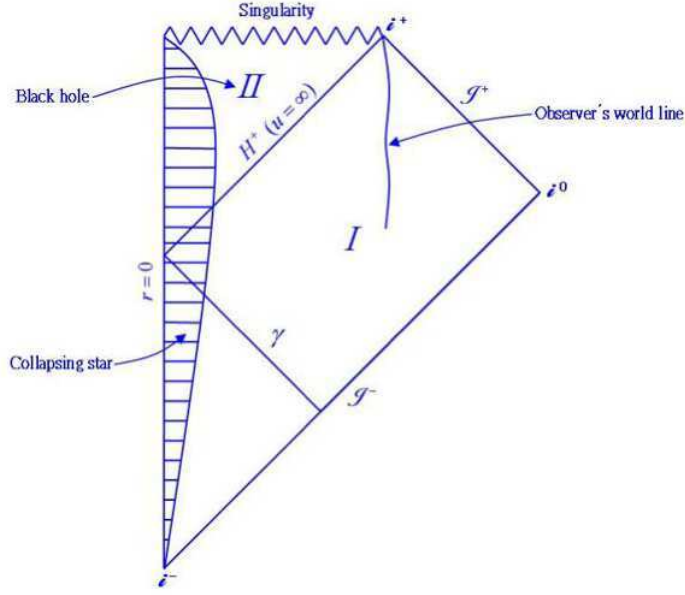}
\caption{Penrose diagram of a collapsing star [14].}
\label{Fig. 1}
\end{figure}

We can estimate the original positive frequency normal mode  on future horizon$(H^+)$:

\begin{eqnarray}
&&A^a_{\mu,s}\propto\varepsilon^a_{\mu,s} \mid {1-\frac{vV}{\left(1-v \right)\left( R_0-2M_{BH}\right)}} \mid^{-i4M_{BH^{\omega}}}\nonumber\\
&&=\begin{cases}
\varepsilon^a_{\mu,s}\left( -1\right) ^{-i4M_{BH^{\omega}}}\left(1-\frac{vV}{\left(1-v \right)\left( R_0-2M_{BH}\right)}\right)^{-i4M_{BH^{\omega}}},~~outside~the~shell\\
\varepsilon^a_{\mu,s}\left(1-\frac{vV}{\left(1-v \right)\left( R_0-2M_{BH}\right)}\right)^{-i4M_{BH^{\omega}}}~~~~~~~~~~~~~~~~~~~,~~inside~the~shell
\end{cases}
\end{eqnarray}
where $\omega$ is the gluon energy.In Eq. (9), we can use the fact that$\left( -1\right) ^{-i4M_{BH^{\omega}}}=e^{4{\pi}M_{BH^{\omega}}}$  . Using equation(9) we observe  that the  states in the horizon satisfy the following condition:
\begin{equation}
\left(A^a_{\mu,s_{matter}}-e^{4{\pi}M_{BH^{\omega}}}A^a_{\mu,s_{in}}\right)\vert{BH,\mu,a,s}\rangle_{matter\bigotimes{in}} =0
\end{equation}
or equivalently

\begin{equation}
\left(A^a_{\mu,s_{in}}-tanh{r_\omega}A^a_{\mu,s_{matter}}\right)\vert{BH,\mu,a,s}\rangle_{matter\bigotimes{in}} =0 
\end{equation}
$tanh{r_\omega}=e^{-4{\pi}M_{BH^{\omega}}}$

which actually constitutes a boundary state. Using the expansion in modes for gluons(equation 1) we may write:

\begin{equation}
\left(\alpha^a_{\mu,s_{in}}-tanh{r_\omega}\alpha^a_{\lambda,s_{matter}^\dagger}\right)\vert{BH,\lambda,a,s}\rangle_{matter\bigotimes{in}} =0 for~n\neq0 
\end{equation}
$tanh{r_\omega}=e^{-4{\pi}M_{BH^{\omega}}}$

Now, we assume that the Kruskal vacuum $\vert{BH,\lambda,a,s\rangle_{in\bigotimes{out}}}$ is related to the Schwarzschild vacuum $\vert{0}\rangle_s$ by

\begin{eqnarray}
\vert{BH,\lambda,a,s}\rangle_{matter\otimes{in}}=F\left(\alpha^a_{\mu,s_{in}},{\alpha}^a_{\lambda,s_{matter}}\right) \vert{0}\rangle_{s}
\end{eqnarray}

where F is some function to be determined later.
From $\left[\alpha^a_{\lambda,s_{in}},\alpha^a_{\lambda,s_{in}^\dagger}\right]=1$,
we obtain $\left[\alpha^a_{\lambda,s_{in}},\left(\alpha^a_{\lambda,s_{in}^\dagger}\right)^m\right]=\frac{\partial}{\partial{\alpha^a_{\lambda,s_{in}^\dagger}}}\left(\alpha^a_{\lambda,s_{in}^\dagger}\right)^m $ 
and $\left[\alpha^a_{\lambda,s_{in}},F\right]=\frac{\partial{F}}{\partial{\alpha^a_{\lambda,s_{in}^\dagger}}}$.

Then using equations (12) and (13), we get the following differential equation for F.
\begin{equation}
\left(\frac{\partial{F}}{\partial{\alpha^a_{\lambda,s_{in}}}^\dagger}-tanh{r_{{{\omega}_n},\lambda}}{\alpha^{a}_{\lambda,s_{matter}}}^{\dagger}F\right)=0
\end{equation}

and the solution is given by
\begin{equation}
F=e^{tanhr_{\omega}{\alpha^a_{\lambda,s_{in}}}^\dagger{\alpha^a_{\lambda,s_{matter}}}^\dagger}
\end{equation}

By substituting (15) into (13)and by properly normalizing the state vector,we get 
\begin{eqnarray}
\vert{BH,\lambda,a,s}\rangle_{{in}\otimes{matter}}=e^{tanhr_{\omega}{\alpha^a_{\lambda,s_{in}}}^\dagger{\alpha^a_{\lambda,s_{matter}}}^\dagger}\vert0\rangle_s=\frac{1}{cosh{r_\omega}}\sum_{n}{tanh^nr_\omega}\vert{n,\lambda,a,s}\rangle_{in}\otimes\vert{n,\lambda,a,s}\rangle_{matter}
\end{eqnarray}
Summing over transversal(physical) degrees of freedom we obtain
\begin{eqnarray}
\vert{BH,\mu,a,s}\rangle_{{in}\otimes{matter}}=\sum_{\lambda=1}^{3}\varepsilon_{\mu}^{\lambda}\vert{BH,\lambda,a,s}\rangle_{{in}\otimes{matter}}=\frac{1}{cosh{r_\omega}}\sum_{n}{tanh^nr_\omega}\vert{n,\mu,a,s}\rangle_{in}\otimes\vert{n,\mu,a,s}\rangle_{matter}
\end{eqnarray}

where $\vert{n,\mu,a,s}\rangle_{in}$  and  $\vert{n,\mu,a,s}\rangle_{matter}$ are orthonormal bases (normal mode solutions) for $H_{in}$ and $H_{matter}$ respectively. We observe that the ground state for gluons inside the black hole horizon  is a maximally entangled with two-mode squeezed states on inside and Matter Hilbert spaces of  black holes. Equation (17) shows that different number of gluons,(n), produced with different probabilities inside and outside the shell  of mini black holes at LHC. These probabilities are related to black hole mass and the energy of gluon $P_{n,\omega}\approx\mid_{in}\langle{n,\mu,a,s}\vert\otimes_{matter}\langle{n,\mu,a,s}\vert{BH,\mu,a,s}\rangle_{{in}\otimes{matter}}\mid^2=\frac{e^{-8{\pi}nM_{BH^{\omega}}}}{cosh^2r_\omega}$ . It seems that many gluons produce near event horizon due to variety in their energy and their probabilities.   We derive the thermal distribution for these groups of gluons outside the shell as the following:

\begin{eqnarray}
N^{gluon}_{\omega,color,spin,in} & = & _{{in}\otimes{matter}}\langle{BH,\mu,a,s}\vert{{\alpha^a_{\mu,s_{in}}}^\dagger\alpha^a_{\mu,s_{in}}}\vert{BH,\mu,a,s}\rangle_{{in}\otimes{matter}} \nonumber\\
& = & _{matter}\langle{n,\mu,a,s}\vert_{in}\langle{n,\mu,a,s}\vert{\frac{1}{cosh^2r_\omega}{\alpha^a_{\mu,s_{in}}}^\dagger\alpha^a_{\mu,s_{in}}}\sum_{n=0}^{\infty}tanh^{2n}\left(r_\omega\right)\vert{n,\mu,a,s}\rangle_{in}\vert{n,\mu,a,s}\rangle_{matter} \nonumber\\
& = & _{matter}\langle{n,\mu,a,s}\vert_{in}\langle{n-1,\mu,a,s}\vert{\frac{1}{cosh^2r_\omega}}\sum_{n=0}^{\infty}tanh^{2n}\left(r_\omega\right)\left(n\right)\vert{n-1,\mu,a,s}\rangle_{in}\vert{n,\mu,a,s}\rangle_{matter} \nonumber\\
& = & \frac{1}{cosh^2r_\omega}\sum_{n=0}^{\omega}e^{-8{\pi}Mn\omega}\left(n\right)\nonumber\\
& = & \frac{1}{cosh^2r_\omega}\frac{e^{-8{\pi}M\omega}}{\left(1-e^{-8{\pi}M\omega}\right)^2}\nonumber\\
& = & \frac{e^{-8{\pi}M\omega}}{1-e^{-8{\pi}M\omega}}
\end{eqnarray}

where ${\alpha^a_{\mu,s_{in}}}^\dagger\alpha^a_{\mu,s_{in}}$ are creation and annihilation operators that act on black hole inside states of gluons. $N^{gluon}_{\omega,color,spin,in}$ is the thermal distribution for gluons with energy $\omega$  and  with a one special color and   spin outside the shell. The thermal distribution for gluons inside the shell can be obtained as following:

\begin{eqnarray}
&& N^{gluon}_{\omega,color,spin,m,matter} \nonumber\\ && = ~_{{in}\otimes{matter}}\langle{BH,\mu,a,s}\vert{{\alpha^a_{\mu,s_{matter}}}^\dagger\alpha^a_{\mu,s_{matter}}}\vert{BH,\mu,a,s}\rangle_{{in}\otimes{matter}} \nonumber\\
&& =  ~_{matter}\langle{n,\mu,a,s}\vert_{in}\langle{n,\mu,a,s}\vert{\frac{1}{cosh^2r_\omega}{\alpha^a_{\mu,s_{matter}}}^\dagger\alpha^a_{\mu,s_{matter}}}\sum_{n=0}^{\infty}tanh^{2n}\left(r_\omega\right)\vert{n,\mu,a,s}\rangle_{in}\vert{n,\mu,a,s}\rangle_{matter} \nonumber\\
&& =  ~_{matter}\langle{n-1,\mu,a,s}\vert_{in}\langle{n,\mu,a,s}\vert{\frac{1}{cosh^2r_\omega}}\sum_{n=0}^{\infty}tanh^{2n}\left(r_\omega\right)\left(n\right)\vert{n-1,\mu,a,s}\rangle_{in}\vert{n,\mu,a,s}\rangle_{matter} \nonumber\\
&& =  \frac{1}{cosh^2r_{\omega}}\sum_{n=0}^{\omega}e^{-8{\pi}mn{\omega}}\left(n\right)\nonumber\\
&& =  \frac{1}{cosh^2r_\omega}\frac{e^{-8{\pi}m\omega}}{\left(1-e^{-8{\pi}m\omega}\right)^2}\nonumber\\
&& =  \frac{e^{-8{\pi}m\omega}}{1-e^{-8{\pi}m\omega}}
\end{eqnarray}

where ${\alpha^a_{\mu,s_{matter}}}^\dagger\alpha^a_{\mu,s_{matter}}$  are creation and annihilation operators that act on black hole matter states of gluons.$N^{gluon}_{\omega,color,spin,m,matter}$ is the thermal distribution for gluons with energy   and  with a one special color and spin inside the shell.m is the fraction of black hole mass that causes to gluon production.
Now we study quark production via Unruh effect inside mini black holes at LHC.
The quark equation in black hole space-time can be calculated as:
\begin{equation}
\left[i\gamma^{\mu}\left(\partial_{\mu}+\Gamma_{\mu}\right)-m_q\right]\psi^a_{q,s}=0
\end{equation}

 Where $a=1,2,3$ is related to the color of quark, s denotes the quark spin, $m_q$ is the quark mass and $q=u,t,b,s,c,d$ represents the quark flavors.The affine connection is given by:
\begin{equation}
\Gamma_{\mu}=-\frac{1}{4}\gamma_{\nu}\left(\partial_{\mu}\gamma^{\nu}+\Gamma^{\nu}_{\mu\lambda}\gamma^{\lambda}\right)
\end{equation}

The gamma matrices are defined by $\gamma_{\mu}=e^a_{\mu}\overline{\gamma}_a$ where $e^a_{\mu}$  are tetrads,$\overline{\gamma}_a$  are the gamma matrices for the inertial frame [18]and the Levi-Civita connection coefficients $\Gamma^{\nu}_{\mu\lambda}$ can be calculated by the Lagrange method[13].Using the Kruskal coordinate we obtain the following solution for equation(20):

\begin{equation}
\psi^a_{q,s_z}\propto
\begin{cases}
u_{rq}^{+a}\left(-1\right)^{-i4M_{BH^{\omega}}}\left(1-\frac{vV}{\left(1-v\right)\left(R_0-2M_{BH}\right)}\right)^{-i4M_{BH^{\omega}}} ~~~~~~~~~~~~~~~~ outside~the~shell
\\v_{r\overline{q}}^{-\overline{a}}\left(1-\frac{vV}{\left(1-v\right)\left(R_0-2M_{BH}\right)}\right)^{-i4M_{BH^{\omega}}} ~~~~~~~~~~~~~~~~~~~~~~~~~~~~~~~~~~~inside~the~shell
\end{cases}
\end{equation}

Where $u_{rq}^{+a}$ and $v_{r\overline{q}}^{-\overline{a}}$ denote the quark and antiquark spin wave functions respectively.Following the calculations in ref[13] we can obtain 
\begin{equation}
\vert{BH,q,a,s}\rangle_{{in}\otimes{matter}}=cosr_{\omega}\sum_{n=0,1}tan^{n}r_{\omega}\vert{n,q,a,s}\rangle_{in}\otimes\vert{n,\overline{q},\overline{a},s }\rangle_{matter}
\end{equation}
$tan\left(r_{\omega}\right)=e^{-4{\pi}M_{BH^{\omega}}},   cos\left(r_{\omega}\right)=\left(1+e^{-8{\pi}M_{BH^{\omega}}}\right)^\frac{-1}{2}$

where $\vert{n,q,a,s}\rangle_{in}~,~~~\vert{n,\overline{q},\overline{a},s }\rangle_{matter}$  , are quark and  antiquarkstates inside and matter Hilbert spaces of black hole. The thermal distribution for this quark production inside the shell is:

\begin{eqnarray}
N_{\omega,color,s,in}^{quark} & = &  _{{in}\otimes{matter}}\langle{BH,q,a,s}\vert{{c^{a\dagger}_{q,s,in}}c^a_{q,s,in}}\vert{BH,q,a,s}\rangle_{{in}\otimes{matter}} \nonumber\\
& = & _{in}\langle{0,\overline{q},\overline{a},s}\vert_{matter}\langle{1,q,a,s}\vert{sin^2\left(r_{\omega}\right)}\vert{1,q,a,s}\rangle_{matter}\vert{0,\overline{q},\overline{a},s}\rangle_{in}\nonumber\\
& = & sin^2\left(r_{\omega}\right)\nonumber\\
& = & \frac{e^{-8{\pi}M_{BH^{\omega}}}}{1+e^{-8{\pi}M_{BH^{\omega}}}}
\end{eqnarray}
\\
Where ${{c^{a\dagger}_{q,s,in}}~,~c^a_{q,s,in}}$  are creation and annihilation operators that act on black hole inside states of quarks.$N_{\omega,color,s,in}^{quark}$ is the thermal distribution for quarks with energy $\omega $  and a special color and spin inside the shell. The thermal distribution for quarks inside the shell can be calculated as following:

\begin{eqnarray}
N_{\omega,color,s,matter}^{quark} & = &  _{{in}\otimes{matter}}\langle{BH,q,a,s}\vert{{c^{a\dagger}_{q,s,matter}}c^a_{q,s,matter}}\vert{BH,q,a,s}\rangle_{{in}\otimes{matter}} \nonumber\\
& = & _{in}\langle{1,\overline{q},\overline{a},s}\vert_{matter}\langle{0,q,a,s}\vert{sin^2\left(r_{\omega}\right)}\vert{0,q,a,s}\rangle_{matter}\vert{1,\overline{q},\overline{a},s}\rangle_{in}\nonumber\\
& = & sin^2\left(r_{\omega}\right)\nonumber\\
& = & \frac{e^{-8{\pi}m\omega}}{1+e^{-8{\pi}m\omega}}
\end{eqnarray}

Where ${{c^{a\dagger}_{q,s,matter}}~,~c^a_{q,s,matter}}$  are creation and annihilation operators that act on black hole matter states of quarks.$N_{\omega,color,s,matter}^{quark}$ is the thermal distribution for quarks with energy $\omega$  and a special color and spin inside the shell.m is the fraction of black hole mass that causes to quark production.

\section{Higgs production inside the $TeV$ black holes}

 We denote the process of proton-proton collision at $TeV$ centre of mass energy forming a colorful black hole. Generically speaking, quantum black holes $(QBHs)$ form representations of $SU(3)_c$ and carry a QED charge. The process of two partons $p_i$, $p_j$ forms a quantum black hole in the c representation of $SU(3)_c$ and charge $q({p_i+p_j}\rightarrow{{QBH}^q_c})$. The following different transitions are possible at a proton collider [15]:
\begin{eqnarray}
\left( a\right)~~3\otimes{\overline{3}} & = & 8\oplus{1}\nonumber\\
\left( b\right)~~3\otimes{3} & = & 6\oplus{\overline{3}}\nonumber\\
\left( c\right)~~3\otimes{8} & = & 3\oplus{\overline{6}}\oplus{15}\nonumber\\
\left( d\right)~~8\otimes{8} & = & 1_s\oplus{8_s}\oplus{8_A}\oplus{\overline{10}_A}\oplus{27_s}\nonumber\\
\left( e\right)~~\overline{3}\otimes{\overline{3}} & = & \overline{6}\oplus{3}\nonumber\\
\left( f\right)~~\overline{3}\otimes{8} & = & \overline{3}\oplus{6}\oplus{\overline{15}}
\end{eqnarray}

The branching fractions for each quantum black hole state has been determined previously [16].These calculations show that quantum black holes decay to a large amount of quarks and gluons .In fact these black holes are expected to be a quark and a gluon factory[2].These quarks and gluons interact with each other and form colorless hadrons or Higgs bosons. We are treating the quarks and gluons as free particles near mini black holes at LHC, because their energy is very high and as also mentioned in our previous section, quark and antiquark are produced inside and outside of event horizon and don$’$t access to each other .Thus we don$’$t expect them to hadronize before forming Higgs boson. Also these calculations show that some Higgs bosons will be radiated directly from mini black holes at LHC.Our aim is to study the effect of Higgs boson production on hadronic cross section .For this reason we consider the process that first, Higgs boson produce via heavy quark and high energy gluon interactions with a subsequent decay to light QCD matter.To obtain the total gluon cross section inside the shell of black holes produced in proton-proton collision we need to multiply the black hole production cross section by the number of gluons produced inside the shell.
\begin{equation}
\sigma_{pp\rightarrow{g},inside~the~shell}=N_{\omega,color,spin,inside~the~shell}^{gluon}\sigma_{pp\rightarrow{BH}}
\end{equation}
And to calculate the total gluon cross section outside the shell,we need to multiply the black hole production cross section by the number of gluons produced outside the shell.
\begin{equation}
\sigma_{pp\rightarrow{g},outside~the~shell}=N_{\omega,color,spin,outside~the~shell}^{gluon}\sigma_{pp\rightarrow{BH}}
\end{equation}
In which [17,18]

\begin{eqnarray}
\sigma^{pp\rightarrow{BH}} & = & \sum_{ab}\int_{\tau}^{1}dx_a\int_{\frac{\tau}{x_a}}^{1}dx_bf_{a/A}\left(x_a,\mu^{2}\right)\times{f_{b/B}\left(x_b,\mu^{2}\right)\sigma^{ab\rightarrow{BH}}\left(\widehat{s}\right)\delta\left(x_ax_b-\frac{M^2_{BH}}{s}\right)}\nonumber\\
\sigma^{ab\rightarrow{BH}} & = & \frac{1}{M_p^2}\left[ \frac{M_{BH}}{M_p}\left(\frac{8\Gamma\left(\frac{d+3}{2}\right)}{d+2}\right)\right]^\frac{2}{d+1}
\end{eqnarray}

where a,b are the interacting partons of two protons A and B, $f_{a/A}$ and $f_{b/B}$ are the parton distribution functions; $x_a$ , $x_b$ are the longitudinal momentum fractions of the partons inside the proton ,$\sqrt{s}$  is the center of mass energy; $M_{BH}$ ,$M_p$ are the black hole mass and plank mass respectively , d is the number of extra dimensions, $\widehat{s}=x_ax_bs=M_{BH}^2,~ \mu^2=M_{BH}^2$  and $\tau=\frac{M_{BH}^2}{s},\sum_{ab}$ represents the sum over all partonic contributions.
These gluons can interact with each other and produce the Higgs boson. The rate for this interaction is[19]: 
\begin{equation}
\Gamma{\left(gg\rightarrow{H}\right)}=\left(\frac{\alpha_s}{\pi}\right)^2\frac{\pi}{288\sqrt{2}}\left[\frac{3}{2\Lambda}\left(1+\left(1-\frac{1}{\Lambda}\right)arcsin^2\left(\sqrt{\Lambda}\right)\right)\right]^2
\end{equation}
\\
in which $\alpha_s$ is the strong coupling constant,$\Lambda=\frac{M_H^2}{4M_t^2}$ is the renormalization scale, $M_H$ is the  Higgs boson mass and $M_t$is the top quark mass.
Electro weak precision data collected at the CERN Large Electron-Positron Collider (LEP),at the Stanford Linear Collider (SLC) and at the Tevatron at Fermilab predict a light SM Higgs boson below approximately 200 GeV. On the other hand, the direct search at LEP2 has excluded Higgs bosons below 114 GeV which leaves a relatively narrow window for the mass[20]. It might be heavy enough to be created by a boson fusion process. However colorful black holes at LHC radiate a large amount of QCD matter and less amount of other fermions and bosons.  This leads to the production of Higgs bosons via quark and gluon interactions. For this reason we discuss Higgs boson production via QCD matter interactions and give out other channels for its production.\\
\indent First, two proton annihilate and black hole is created , then black hole decays to gluons; next  gluons interact with each other and Higgs boson is produced. We assume gluons annihilate and produce heavy quark - antiquarks with a mass $M_q$ and a subsequent decay to Higgs boson. Here we assume that these heavy quarks are top quarks. Since the top quark mass $(M_t =175GeV)$ is close to Higgs boson mass, these quarks could easily interact with each other and producing Higgs bosons. We denote the Higgs boson production cross section outside the shell of black hole by $\sigma_{H,out~side~shell}^1$   and calculate   $\sigma_{H,out~side~shell}^1$   as the following:
  
\begin{eqnarray}
&&\sigma_{H,out~side~shell}^1\left(pp\rightarrow{H}\right)\nonumber\\
&& = f\times{\sum_{ab}\int_\tau^1dx_a\int_\frac{\tau}{x_a}^1dx_b\int\frac{d\omega_1}{2{\pi}}\int\frac{d\omega_2}{2{\pi}}\int{dz\delta\left(x_ax_b-\frac{M_{BH}^2}{s}\right)2{\pi}\delta\left(zM_{BH}-\omega_1-\omega_2\right)2{\pi}\delta\left(M_H-\omega_1-\omega_2\right)}}\nonumber\\
&& ~~~~~ \times{f_{a/A}\left(x_a,\mu^2\right)f_{b/B}\left(x_b,\mu^2\right)L^{ab\rightarrow{BH}\rightarrow{gg}}\left(\omega_1,\omega_2\right)p_g\left(\omega_1\right)p_g\left( \omega_2\right) L^{gg\rightarrow{H}}\left(\omega_1,\omega_2\right)}\nonumber\\
&& = f\times \sum_{ab}\int_\tau^1dx_a\int_\frac{\tau}{x_a}^1dx_b\int\frac{d\omega_1}{2{\pi}}\int\frac{d\omega_2}{2{\pi}}\int{dz{\delta}\left(x_ax_b-\frac{M_{BH}^2}{s}\right)2{\pi}\delta\left(zM_{BH}-\omega_1-\omega_2\right)2{\pi}\delta\left(M_H-\omega_1-\omega_2\right)}\nonumber\\
&& ~~~~~ \times f_{a/A}\left(x_a,\mu^2\right)f_{b/B}\left(x_b,\mu^2\right)\frac{e^{-8{\pi}M_{BH^{\omega_1}}}}{1-e^{-8{\pi}M_{BH^{\omega_1}}}}\frac{e^{-8{\pi}M_{BH^{\omega_2}}}}{1-e^{-8{\pi}M_{BH^{\omega_2}}}}\frac{1}{M_p^2}\left[ \frac{\omega_1+\omega_2}{zM_p}\left( \frac{4\Gamma\left(\frac{7}{2}\right)}{3}\right)\right]^\frac{2}{5}\frac{1}{\omega_1^2}\frac{1}{\omega_2^2}\nonumber\\
&& ~~~~~ \times \sum_{c=1}^{8}\left( T^cT^c\right)_{mn}\left(\frac{\alpha_s}{\pi} \right)^2\frac{\pi}{288\sqrt{2}}\left[ \frac{6M_t^2}{\left( \omega_1+\omega_2\right)^2 }\left( 1+\left( 1- \frac{4M_t^2}{\left( \omega_1+\omega_2\right)^2}\right)arcsin^2\left(\sqrt{\frac{\left(\omega_1+\omega_2 \right)^2 }{4M_t^2}}\right)\right)\right]^2\nonumber\\
&& \approx f\times C_F\delta_{mn}\left(\frac{s}{M_{BH}^2}\right)^{1.2}\left[1+3\frac{M_{BH}^2}{s}ln\left(\frac{s}{M_{Bh}^2}\right)\right]\left(\frac{\alpha_s}{\pi}\right)^2\frac{\pi}{288\sqrt{2}} \nonumber\\
&& ~~~~~ \times \left[\frac{6M_t^2}{M_H^2}\left(1+\left(1-\frac{4M_t^2}{M_H^2}\right)arcsin^2\left(\sqrt{\frac{M_H^2}{4M_t^2}}\right)\right)\right]^2 \frac{1}{M_P^2}\left[\frac{M_{BH}}{M_P}\left(\frac{4\Gamma\left(\frac{7}{2}\right)}{3}\right)\right]^\frac{2}{5} \nonumber\\
&& ~~~~~ \times \frac{e^{-8{\pi}M_{BH}^2}}{\left(1-e^{-8{\pi}M_{BH}^2}\right)}\left[\frac{1}{M_t^2}-\frac{1}{M_H^2}+ln\left(\frac{M_t}{M_H}\right)\right]
\end{eqnarray}
\\
where $f$ is the fraction of Higgs boson that transforms to outside of event horizon.For High energy Higgs bosons f will be unity[2],However for low energy Higgs boson,f will be at least 0.85[21].We choose $f=0.85$in this research. Also $z$ is the fraction of black hole mass that transforms to gluon energy .When black hole decays to gluons,each of these particles carry a fraction of black hole mass. This means that black hole mass decreases with each gluon emitted and we should integrate over $z$. We also define the gluon part, Higgs part and gluon propagator for this integral as: 

\begin{equation}
L^{ab\rightarrow{BH}\rightarrow{gg}}\left(\omega_1,\omega_2\right)=\frac{e^{-8{\pi}M_{BH^{\omega_1}}}}{\left(1-e^{-8{\pi}M_{BH^{\omega_1}}}\right)}\frac{e^{-8{\pi}M_{BH^{\omega_2}}}}{\left(1-e^{-8{\pi}M_{BH^{\omega_2}}}\right)}\frac{1}{M_P^2}\left[\frac{\omega_1+\omega_2}{zM_P}\left(\frac{4\Gamma\left(\frac{7}{2}\right)}{3}\right)\right]^\frac{2}{5}
\end{equation}

\begin{eqnarray}
&& L^{gg\rightarrow{H}}\left(\omega_1,\omega_2\right)\nonumber\\
&& = \sum_{C=1}^{8}\left(T^CT^C\right)_{mn}\left(\frac{\alpha}{s{\pi}}\right)^2\frac{\pi}{288\sqrt{2}}\left[\frac{6M_t^2}{\left(\omega_1+\omega_2\right)^2}\left( 1+\left( 1- \frac{4M_q^2}{\left( \omega_1+\omega_2\right)^2}\right)arcsin^2\left(\sqrt{\frac{\left(\omega_1+\omega_2 \right)^2 }{4M_t^2}}\right)\right)\right]^2\nonumber\\
&& = C_F\delta_{mn}\left(\frac{\alpha_s}{\pi} \right)^2\frac{\pi}{288\sqrt{2}}\left[ \frac{6M_t^2}{\left( \omega_1+\omega_2\right)^2 }\left( 1+\left( 1- \frac{4M_t^2}{\left( \omega_1+\omega_2\right)^2}\right)arcsin^2\left(\sqrt{\frac{\left(\omega_1+\omega_2 \right)^2 }{4M_t^2}}\right)\right)\right]^2
\end{eqnarray}

\begin{equation}
C_F=\sum_{C=1}^{8}\left(T^CT^C\right)_{mn}=\frac{4}{3}\delta_{mn} ~~~~~~ for~~ SU(3)
\end{equation}

\begin{equation}
p_g\left(\omega_1\right)=\frac{1}{\omega_1^2}
\end{equation}

where $C_F$ is the color factor ,$T^C$’s are color group generators and m,n are  the color indices for gluons. Two gluons that come out of black hole should have the correct color to form a color singlet and hence Higgs boson.For this reason we add one suppression color factor $(C_F)$ to equation(23).

\begin{figure}
\includegraphics[scale=1.3]{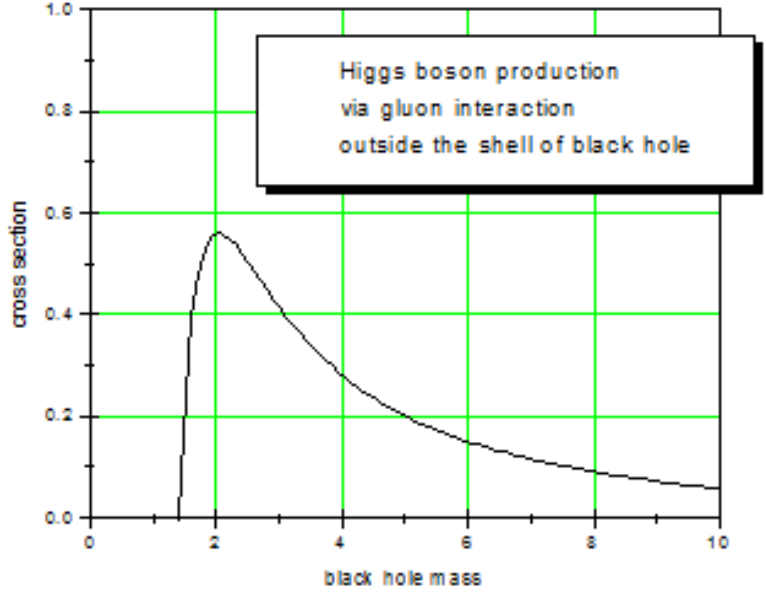}
\caption{The Higgs boson production cross section$(\sigma(pp\rightarrow{BH}\rightarrow{gg}\rightarrow{Higgs}))$ outside the shell as a function of black hole mass.}
\label{Fig. 2}
\end{figure}

In Fig.2 we present the Higgs boson production cross section as a function of black hole mass. In this plot we choose $M_t =175GeV,M_H=130GeV,M_P=2TeV$  and  $\sqrt{s}=14TeV$ for top quark mass, Higgs boson mass, Planck mass and center of mass energy respectively.It is clear that the cross section of Higgs boson produced via gluon fusion outside the shell of black holeis much larger for smaller black hole mass. This is because as the mass of the black hole becomes smaller the temperature becomes larger and the thermal radiation of gluons is enhanced.
With similar calculations to outside the shell,we can calculate the production cross section inside the shell$(\sigma^1_{H,inside~the~shell})$:

\begin{eqnarray}
&&\sigma^1_{H,inside~the~shell}\left(pp\rightarrow{H}\right)\nonumber\\
&& \approx f\times \int_{0}^{M_{BH}}dmC_F\delta_{mn}\left(\frac{s}{m^2}\right)^{1.2}\left[1+3\frac{m^2}{s}ln\left(\frac{s}{m^2}\right)\right]\left(\frac{\alpha_s}{\pi}\right)^2\frac{\pi}{288\sqrt{2}} \nonumber\\
&& ~~~~~ \times  \left[\frac{6M_t^2}{M_H^2}\left(1+\left(1-\frac{4M_t^2}{M_H^2}\right)arcsin^2\left(\sqrt{\frac{M_H^2}{4M_t^2}}\right)\right)\right]^2 \frac{1}{M_P^2}\left[\frac{m}{M_P}\left(\frac{4\Gamma\left(\frac{7}{2}\right)}{3}\right)\right]^\frac{2}{5} \nonumber\\
&& ~~~~~ \times \frac{e^{-8{\pi}m^2}}{\left(1-e^{-8{\pi}m^2}\right)}\left[\frac{1}{M_t^2}-\frac{1}{M_H^2}+ln\left(\frac{M_t}{M_H}\right)\right]\nonumber\\
&& \approx  f\times{C_F\delta_{mn}\left(8{\pi}\left(\frac{s}{M_{BH}}\right)^{1.2}-\left(\frac{s}{M_{BH}^3}\right)^{1.2}\right)\left[1+3\frac{M_{BH}^2}{s}ln\left(s\right)\frac{M_{BH}^3}{s}ln\left(\frac{s}{M_{BH}^2}\right)+3\frac{M_{BH}^3}{s}\left(1-ln\left(M_{BH}^2\right)\right)\right]}\nonumber\\
&& ~~~~~\times \frac{1}{M_P^2}\left[\frac{M_{BH}}{M_P}\left(\frac{4\Gamma\left(\frac{7}{2}\right)}{3}\right)\right]^{\frac{2}{5}}\left[\frac{1}{M_t^2}-\frac{1}{M_H^2}+ln\left(\frac{M_t}{M_H}\right)\right] \left(\frac{\alpha_s}{\pi}\right)^2\frac{\pi}{288\sqrt{2}} \nonumber\\
&& ~~~~~\times \left[\frac{6M_t^2}{M_H^2}\left(1+\left(1-\frac{4M_t^2}{M_H^2}\right)arcsin^2\left(\sqrt{\frac{M_H^2}{4M_t^2}}\right)\right)\right]^2
\end{eqnarray}

\begin{figure}
\includegraphics[scale=1.3]{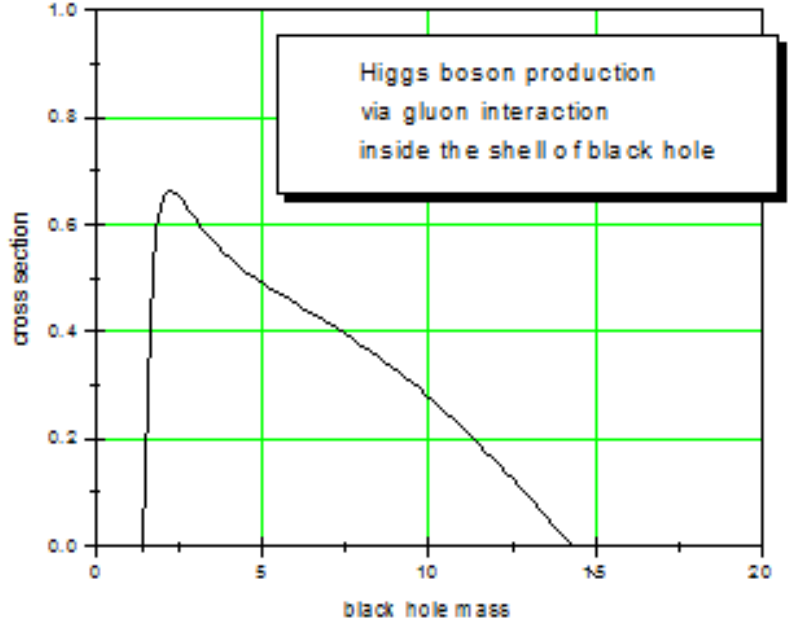}
\caption{The Higgs boson production cross section$(\sigma(pp\rightarrow{BH}\rightarrow{gg}\rightarrow{Higgs}))$ inside the shell as a function of black hole mass.}
\label{Fig. 3}
\end{figure}

In Fig.3 we present the Higgs boson production cross section inside the shell as a function of black hole mass. In this plot we choose $M_t =175GeV,M_H=130GeV ,M_P=2TeV  and \sqrt{s} 14TeV$ for top quark mass, Higgs boson mass, Planck mass and center of mass energy respectively.Comparing this cross section with cross section outside the shell,we find that more gluons produce inside the shell.
    Now we discuss the Higgs boson cross section via quark fusion near mini black holes at LHC. The total quark cross section outside the shell of black holes produced in proton proton collisions is:
 
\begin{equation}
\sigma_{pp\rightarrow{q,outside~the~shell}}=N^{quark}_{\omega,color,spin,outside~the~shell}\sigma_{pp\rightarrow{BH}}
\end{equation}

The bottom quarks produced near event horizon of mini black holes can interact with each other and produce Higgs boson. We only consider low mass Higgs boson production due to bottom quark interaction. The rate for this interaction is [22]:
\begin{equation}
\Gamma^{b\overline{b}\rightarrow{H}}=\frac{M_b^2\alpha_s}{6{\nu^2}M_H^2}\left[-4C_Fln\left(1-\frac{M_H^2}{s}\right)+2C_Fln\left(\frac{M_b^2}{M_H^2}\right)+2C_F\right]
\end{equation}
where $\nu=246GeV, C_F=\frac{4}{3}$

We denote the Higgs boson production cross section via process ${pp\rightarrow{BH}\rightarrow{b\overline{b}}\rightarrow{H}}$ outside the shell by $\sigma_H^2$ with the following calculations:

\begin{eqnarray}
&& \sigma_{H,out~side~shell}^2\left(pp\rightarrow{H}\right)\nonumber\\
&& = f\times{\sum_{ab}\int_\tau^1dx_a\int_\frac{\tau}{x_a}^1dx_b\int\frac{d\omega_1}{2{\pi}}\int\frac{d\omega_2}{2{\pi}}\int{dz\delta\left(x_ax_b-\frac{M_{BH}^2}{s}\right)2{\pi}\delta\left(zM_{BH}-\omega_1-\omega_2\right)2{\pi}\delta\left(M_H-\omega_1-\omega_2\right)}}\nonumber\\
&& \times{f_{a/A}\left(x_a,\mu^2\right)f_{b/B}\left(x_b,\mu^2\right)L^{ab\rightarrow{BH}\rightarrow{q\overline{q}}}\left(\omega_1,\omega_2\right)p_q\left(\omega_1\right)p_{\overline{q}}\left( \omega_2\right) L^{q\overline{q}\rightarrow{H}}\left(\omega_1,\omega_2\right)}\nonumber\\
&& = f\times{\sum_{ab}\int_\tau^1dx_a\int_\frac{\tau}{x_a}^1dx_b\int\frac{d\omega_1}{2{\pi}}\int\frac{d\omega_2}{2{\pi}}\int{dz{\delta}\left(x_ax_b-\frac{M_{BH}^2}{s}\right)2{\pi}\delta\left(zM_{BH}-\omega_1-\omega_2\right)2{\pi}\delta\left(M_H-\omega_1-\omega_2\right)}}\nonumber\\
&& \times{f_{a/A}}\left(x_a,\mu^2\right)f_{b/B}\left(x_b,\mu^2\right)\frac{e^{-8{\pi}M_{BH^{\omega_1}}}}{1+e^{-8{\pi}M_{BH^{\omega_1}}}}\frac{e^{-8{\pi}M_{BH^{\omega_2}}}}{1+e^{-8{\pi}M_{BH^{\omega_2}}}}\frac{1}{M_p^2}\left[ \frac{\omega_1+\omega_2}{zM_p}\left( \frac{4\Gamma\left(\frac{7}{2}\right)}{3}\right)\right]^\frac{2}{5}\frac{\gamma_{\mu}\omega_1^{\mu}+M_b}{\omega_1^2-M_b^2}\nonumber\\
&& \times\frac{\gamma_{\nu}{\omega}_2^{\nu}+M_b}{{\omega}_2^2-M_b^2}N_{color}\delta_{cd}\frac{\left(\omega_1+\omega_2\right)^2\alpha_s}{24{\nu}^2M_H^2}\left[-4C_Fln\left(1-\frac{x_ax_bM_H^2}{M_{BH}^2}\right)+2C_Fln\left(\frac{\left(\omega_1+\omega_2\right)^2}{4M_H^2}\right)+2C_F\right]\nonumber\\
&& \approx N_{color}\delta_{cd}\left(\frac{s}{M_{BH}^2}\right)^{1.2}\left[1+3\frac{M_{BH}^2}{s}ln\left(\frac{s}{M_{BH}^2}\right)\right]\frac{M_b^2\alpha_s}{24{\nu}^2M_H^2}\left[-4C_Fln\left(1-\frac{M_H^2}{M_{BH}^2}\right)+2C_Fln\left(\frac{1}{4}\right)+2C_F\right]\nonumber\\
&& \times\frac{1}{M_P^2}\left[\frac{M_{BH}}{M_P}\left(\frac{4\Gamma\left(\frac{7}{2}\right)}{3}\right)\right]^{\frac{2}{5}}\frac{e^{-8{\pi}M_{BH}^2}}{1-e^{-8{\pi}M_{BH}^2}}\left[-ln\left(\frac{M_H-M_b}{M_H+M_b}\right)+\frac{M_H^2}{M_b^2}\right]
\end{eqnarray}

where we define the quark part , Higgs part and quark propagator for this integral as: 
\begin{equation}
L^{ab\rightarrow{BH}\rightarrow{q\overline{q}}}\left(\omega_1,\omega_2\right)=\frac{e^{-8{\pi}M_{BH^{\omega_1}}}}{1+e^{-8{\pi}M_{BH^{\omega_1}}}}\frac{-8{\pi}M_{BH^{\omega_2}}}{1+e^{-8{\pi}M_{BH^{\omega_2}}}}\frac{1}{M_P^2}\left[\frac{\omega_1+\omega_2}{zM_P}\left(\frac{4\Gamma\left(\frac{7}{2}\right)}{3}\right)\right]^{\frac{2}{5}}
\end{equation}

\begin{equation}
L^{q\overline{q}\rightarrow{H}}\left(\omega_1,\omega_2\right)=N_{color}\delta_{cd}\frac{\left(\omega_1+\omega_2\right)^2\alpha_s}{24{\nu}^2M_H^2}\left[-4C_Fln\left(1-\frac{M_H^2}{s}\right)+2C_Fln\left(\frac{\left(\omega_1+\omega_2\right)^2}{4M_H^2}\right)+2C_F\right]
\end{equation}

\begin{equation}
P_q=\frac{\gamma_\mu\omega_1^\mu+M_b}{\omega_1^2-M_b^2}
\end{equation}

$z$ is the fraction of black hole mass that transforms to quark energy, $N_{color}$  is  the the number of colors and $c,d$ are  the color indices for quarks.

\begin{figure}
\includegraphics[scale=1.3]{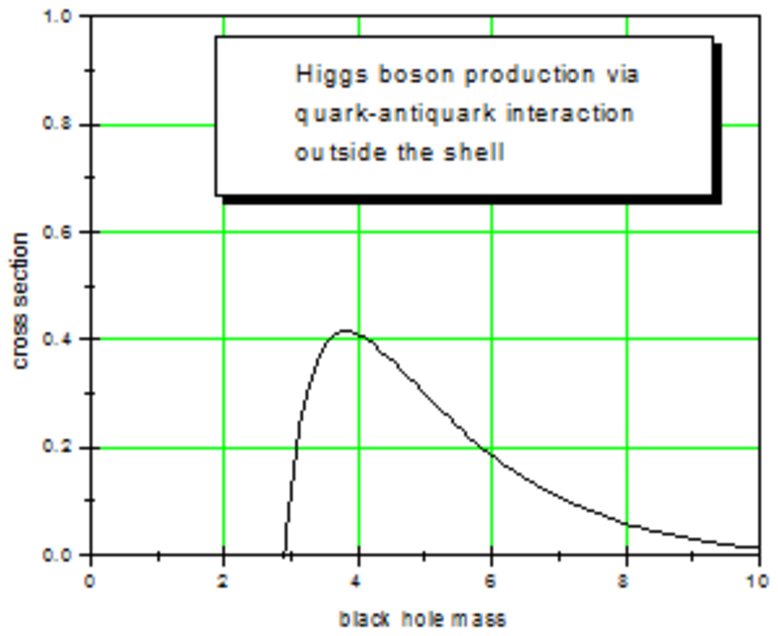}
\caption{The Higgs boson production cross section$(\sigma(pp\rightarrow{BH}\rightarrow{q\overline{q}}\rightarrow{Higgs}))$ outside the shell as a function of black hole mass.}
\label{Fig. 4}
\end{figure}

In Fig. 4 we present the results for the cross section of Higgs boson produced via quark-anti quark interactions outside the shell of mini black holes as a function of black hole mass.In this plot we choose $M_b=5GeV,M_H=130GeV,M_P=2TeV and \sqrt{s}=14TeV$  for bottom quark mass, Higgs boson mass, Planck mass and center of mass energy respectively. As can be seen from the figure 3,the cross sections decreases rapidly when the black hole mass increases.
Using the result for outside the shell,we calculate the production cross section inside the shell $(\sigma^2_{H,inside~the~shell})$:

\begin{eqnarray}
&& \sigma^2_{H,inside~the~shell}\left(pp\rightarrow{H}\right)\nonumber\\
&& \approx f\times \int_{0}^{M_{BH}}dmN_{color}\delta_{cd}\left(\frac{s}{m^2}\right)^{1.2}\left[1+3\frac{m^2}{s}ln\left(\frac{s}{m^2}\right)\right]\frac{M_b^2\alpha_s}{24{\nu}^2M_H^2}\left[-4C_Fln\left(1-\frac{M_H^2}{m^2}\right)+2C_Fln\left(\frac{1}{4}\right)+2C_F\right]\nonumber\\
&& ~~~~~ \times \frac{1}{M_P^2}\left[\frac{m}{M_P}\left(\frac{4\Gamma\left(\frac{7}{2}\right)}{3}\right)\right]^\frac{2}{5}\frac{e^{-8{\pi}m^2}}{1+e^{-8{\pi}m^2}}\left[-ln\left(\frac{M_H-M_b}{M_H+M_B}\right)+\frac{M_H^2}{M_b^2}\right]\nonumber\\
&& \approx f\times N_{color}\delta_{cd}\left[8{\pi}\left(\frac{s}{M_{BH}}\right)^{1.2}+\left(\frac{s}{m_{BH}^3}\right)^{1.2}\right] \nonumber\\  && ~~~~~ \times \left[1+3\frac{M_{BH}^2}{s}ln\left(s\right)+\frac{M_{BH}^3}{s}ln\left(\frac{s}{M_{BH}^2}\right)+3\frac{M_{BH}^3}{s}\left(1-ln\left(M_{BH}^2\right)\right)\right]\nonumber\\
&& ~~~~~ \times\frac{1}{M_P^2}\left[\frac{M_{BH}}{M_P}\left(\frac{4\Gamma\left(\frac{7}{2}\right)}{3}\right)\right]^{\frac{2}{5}}\left[-ln\left(\frac{M_H-M_b}{M_H+M_b}\right)+\frac{M_H^2}{M_b^2}\right]\frac{M_b^2\alpha_s}{24{\nu}^2M_H^2}\nonumber\\
&& ~~~~~ \times\left[-4C_FM_{BH}^2\left(1-ln\left(1-\frac{M_H^2}{M_{BH}^2}\right)\right)+2C_Fln\left(\frac{1}{4}\right)+2C_F\right]
\end{eqnarray}

\begin{figure}
\includegraphics[scale=1.3]{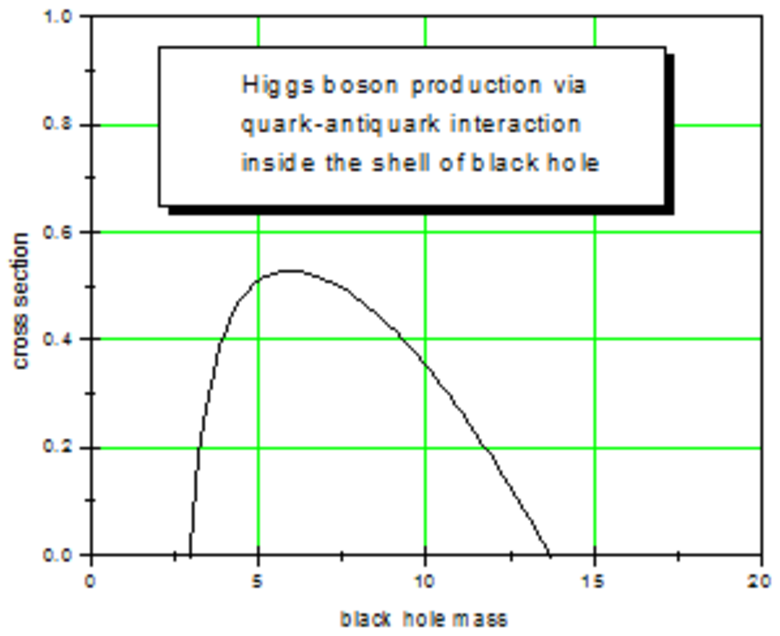}
\caption{The Higgs boson production cross section$(\sigma(pp\rightarrow{BH}\rightarrow{q\overline{q}}\rightarrow{Higgs}))$ inside the shell as a function of black hole mass.}
\label{Fig. 5}
\end{figure}

 In Fig. 5 we present the results for the cross section of Higgs boson produced via quark-antiquark interactions inside the shell of mini black holes as a function of black hole mass.In comparing with previous cross section $(\sigma_{H,outside~the~shell}^2 )$,it is observed that this cross section is more effective at LHC.
Now we calculate the cross section for producing Higgs inside black hole:the black hole vacuum for Higgs will evolve into a state, called the Unruh state, which can be formulated as [10]:

\begin{equation}
\vert{BH}\rangle_{in\otimes{matter}}=\frac{1}{cosh\left(r_{\omega}\right)}\sum{tanh^n\left(r_{\omega}\right)\vert{n}\rangle_{in}\bigotimes\vert{n}\rangle_{matter}}
\end{equation}

where $tanh(r_{\omega})=e^{-4{\pi}M_{BH^{\omega}}},~~ cosh(r_{\omega})=(1-e^{-8{\pi}M_{BH^{\omega}}})^{-\frac{1}{2}}$

For calculating Higgs cross section outside the shell we need to multiply its distribution by black hole production cross section. We write:

\begin{eqnarray}
N^{Higgs}_{\omega,outside~the~shell}& = & _{in\otimes{matter}}\langle{BH}\vert{a^t_{in}a_{in}}\vert{BH}\rangle_{in\otimes{matter}}\nonumber\\
& = & _{matter}\langle{n}\vert_{in}\langle{n-1}\vert{\frac{1}{cosh^2\left(r_{\omega}\right)}}\sum_{n=0}^{\infty}tanh^{2n}\left(r_{\omega}\right)n\vert{n-1}\rangle_{in}\vert{n}\rangle_{matter}\nonumber\\
& = & _{matter}\langle{n}\vert{\frac{1}{cosh^2\left(r_{\omega}\right)}}\sum_{n=0}^{\infty}tanh^{2n+2}\left(r_{\omega}\right)\left(n+1\right)\vert{n}\rangle_{matter}\nonumber\\
& = & \frac{sinh^2\left(r_{\omega}\right)}{cosh^4\left(r_{\omega}\right)}\sum_{n=o}^{\omega}tanh^{2n}\left(r_{\omega}\right)\left(n+1\right)\nonumber\\
& = & \frac{sinh^2\left(r_{\omega}\right)}{cosh^4\left(r_{\omega}\right)}\frac{1}{\left(1-tanh^2\left(r_{\omega}\right)\right)^2}\nonumber\\
& = & sinh^2\left(r_{\omega}\right)\frac{cosh^4\left(r_{\omega}\right)}{cosh^4\left(r_{\omega}\right)}\nonumber\\
& = & sinh^2\left(r_{\omega}\right)\nonumber\\
& = & \frac{e^{-8{\pi}M_{BH^{\omega}}}}{\left(1-e^{-8{\pi}M_{BH^{\omega}}}\right)}
\end{eqnarray}

where $a_{in}^t,a_{in}$ are the creation and annihilation operators that act on outside the shell of black hole.
We derive the Higgs boson production cross section $(\sigma^3H)$ as:

\begin{eqnarray}
&& \sigma^3_{pp\rightarrow{H},outside~the~shell}\nonumber\\
&& = f\times{\sum_{ab}\int_{\tau}^{1}dx_a\int_{\frac{\tau}{x_a}}^{1}dx_b\int\frac{d{\omega}}{2{\pi}}\int{dz\delta\left(x_ax_b-\frac{M_{BH}^2}{s}\right)2{\pi}\delta\left(zM_{BH}-\omega\right)}}\nonumber\\
&&\times{f_{a/A}}\left(x_a,\mu^2\right)f_{b/B}\left(x_b,\mu^2\right)\frac{e^{-8{\pi}M_{BH^{\omega}}}}{\left(1-e^{-8{\pi}M_{BH^{\omega}}}\right)}\frac{1}{M_P^2}\left[\frac{\omega}{zM_P}\left(\frac{4\Gamma\left(\frac{7}{2}\right)}{3}\right)\right]^{\frac{2}{5}}\nonumber\\
&& \approx f\times{\left(\frac{s}{M_{BH}^2}\right)^{1.2}\left[1+3\frac{M_{BH}^2}{s}ln\left(\frac{s}{M_{BH}^2}\right)\right]\frac{1}{M_P^2}\left[\frac{M_{BH}}{M_P}\left(\frac{4\Gamma\left(\frac{7}{2}\right)}{3}\right)\right]^{\frac{2}{5}}}\nonumber\\
&&\times{\frac{1}{8{\pi}M_{BH}}\left[ln\frac{M_{BH}^2}{M_{H}^2}+ln\left(\frac{M_{BH}^2}{M_{H}^2}-1\right)+2ln\left(2\right)\right]}
\end{eqnarray}

\begin{figure}
\includegraphics[scale=1.3]{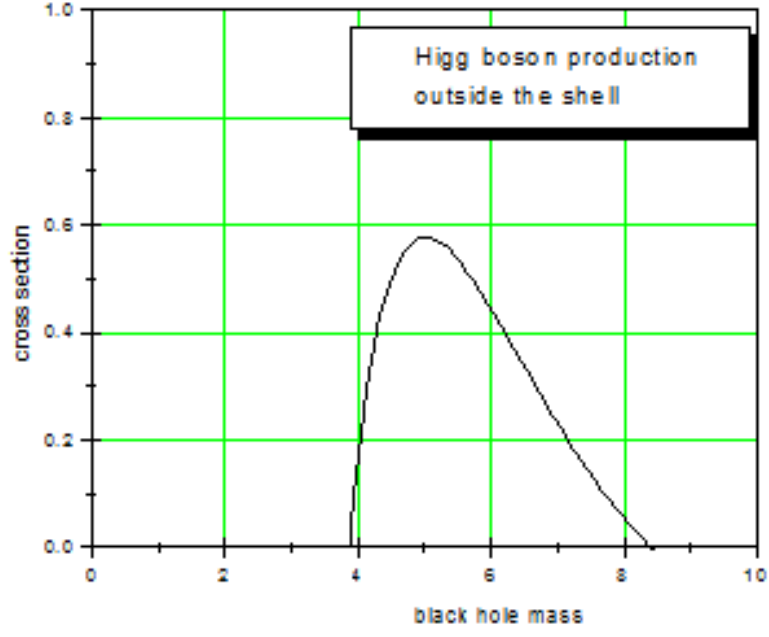}
\caption{The Higgs boson production cross section$(\sigma(pp\rightarrow{BH}rightarrow{Higgs}))$ outside the shell as a function of black hole mass.}
\label{Fig. 6}
\end{figure}

In Fig. 6 we present the cross section forHiggs boson radiated from outside the shell of mini black holes as a function of black hole mass. In this plot we choose $M_H=130GeV,~M_P=2TeV~~ and ~~ \sqrt{s}=14TeV$ for Higgs boson mass, Planck mass and center of mass energy respectively. The Higgs boson production cross section decreases at LHC as the mass of the black hole increases.
To get cross section inside the shell we need to multiply its distribution by black hole production cross section. We obtain:

\begin{eqnarray}
N^{Higgs}_{\omega,outside~the~shell}& = & _{in\otimes{matter}}\langle{BH}\vert{a^t_{matter}a_{matter}}\vert{BH}\rangle_{in\otimes{matter}}\nonumber\\
& = & _{matter}\langle{n-1}\vert_{in}\langle{n}\vert{\frac{1}{cosh^2\left(r_{\omega}\right)}}\sum_{n=0}^{\infty}tanh^{2n}\left(r_{\omega}\right)n\vert{n}\rangle_{in}\vert{n-1}\rangle_{matter}\nonumber\\
& = & _{in}\langle{n}\vert{\frac{1}{cosh^2\left(r_{\omega}\right)}}\sum_{n=0}^{\infty}tanh^{2n+2}\left(r_{\omega}\right)\left(n+1\right)\vert{n}\rangle_{in}\nonumber\\
& = & \frac{sinh^2\left(r_{\omega}\right)}{cosh^4\left(r_{\omega}\right)}\sum_{n=o}^{\omega}tanh^{2n}\left(r_{\omega}\right)\left(n+1\right)\nonumber\\
& = & \frac{sinh^2\left(r_{\omega}\right)}{cosh^4\left(r_{\omega}\right)}\frac{1}{\left(1-tanh^2\left(r_{\omega}\right)\right)^2}\nonumber\\
& = & sinh^2\left(r_{\omega}\right)\frac{cosh^4\left(r_{\omega}\right)}{cosh^4\left(r_{\omega}\right)}\nonumber\\
& = & sinh^2\left(r_{\omega}\right)\nonumber\\
& = & \frac{e^{-8{\pi}M_{BH^{\omega}}}}{\left(1-e^{-8{\pi}M_{BH^{\omega}}}\right)}
\end{eqnarray}

where $a_{matter}^t,a_{matter}$  are the creation and annihilation operators that act on inside the shell of black hole states.
    Using the result in equation(46),We obtain the Higgs boson production cross section  $(\sigma^3_{\omega,inside~the~shell})$ as:

\begin{eqnarray}
&& \sigma^3_{pp\rightarrow{H},inside~the~shell}\nonumber\\
&& \approx  f\times \int_{0}^{M_{BH}}dm\left(\frac{s}{m^2}\right)^{1.2}\left[1+3\frac{m^2}{s}ln\left(\frac{s}{m^2}\right)\right]\frac{1}{M_P^2}\left[\frac{m}{M_P}\left(\frac{4\Gamma\left(\frac{7}{2}\right)}{3}\right)\right]^{\frac{2}{5}} \nonumber\\
&& ~~~~~ \times \frac{1}{8{\pi}m}\left[ln\left(\frac{m^2}{M_H^2}\right)+ln\left(\frac{m^2}{M_H^2}-1\right)+2ln\left(2\right)\right]\nonumber\\
&& \approx f\times\left[8{\pi}\left(\frac{s}{M_{BH}}\right)^{1.2}-\left(\frac{s}{M_{BH}^3}\right)^{1.2}\right]\left[1+3\frac{M_{BH}^3}{s}ln\left(s\right)+3\frac{M_{BH}^3}{s}\left(1-ln\left(M_{BH}^2\right)\right)\right]\frac{1}{M_P^2}\left[\frac{M_{BH}}{M_P}\left(\frac{4\Gamma\left(\frac{7}{2}\right)}{3}\right)\right]^{\frac{2}{5}}\nonumber\\
&& ~~~~~ \times\left[\frac{M_{BH}^2}{M_H^2}\left(ln\left(\frac{M_{BH}^2}{M_H^2}\right)+1\right)+\left(\frac{M_{BH}^2}{M_H^2}-1\right)\left(ln\left(\frac{M_{BH}^2}{M_H^2}-1\right)+1\right)\right]
\end{eqnarray}

\begin{figure}
\includegraphics[scale=1.3]{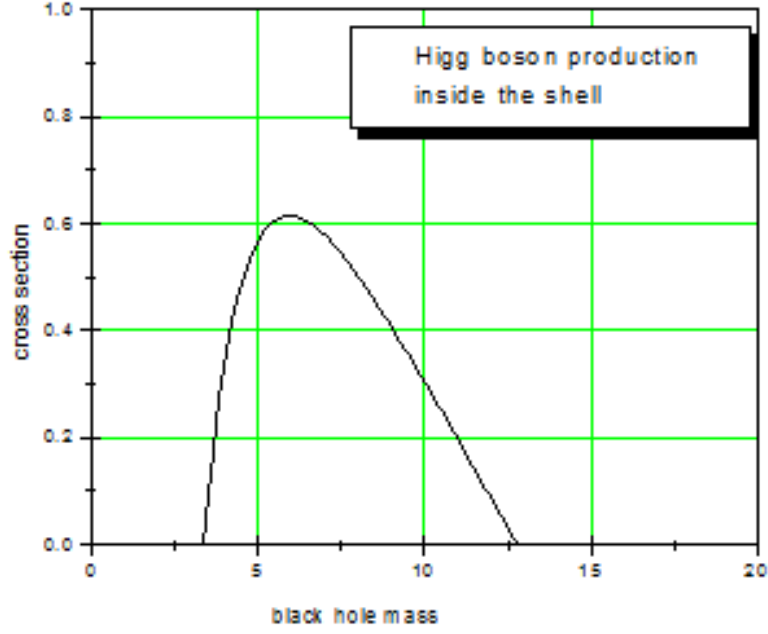}
\caption{The Higgs boson production cross section$(\sigma(pp\rightarrow{BH}rightarrow{Higgs}))$ inside the shell as a function of black hole mass.}
\label{Fig. 7}
\end{figure}

   In Fig. 7 we present the cross section for Higgs boson radiated from inside the shell of mini black holes as a function of black hole mass.Comparing figure (6)with figure (7) ,we observe that at lower mass $"M_{BH}<4TeV"$ and at higher mass $”M_{BH}>9TeV”$, the Higgs boson will not be produced outside the shell of black hole,However this particle can be produced inside the shell for $”2.5 TeV<M_{BH}<12.5TeV”$.Also figures(2,4) show that the outside the shell black hole is able to produce a quark for $”3TeV<M_{BH}<10TeV”$; and eventually for $“M_{BH}<3TeV and M_{BH}>12.5TeV”$ the black hole can only emit massless gluons.However figures(3,5)state that inside the shell of black hole is able to produce a quark for  $”2.5TeV<M_{BH}<14TeV”$.
   
\section{PQCD processes for Higgs production}

In this section we consider different channels for Higgs production in flat space time. Suppose that two protons annihilate and heavy quark and anti heavy quarks with momentum q produce, and then these particles  emit Higgs boson.
To obtain the total Higgs cross section in proton-proton collisions we need to multiply the top quark cross section from proton- proton annihilation by the $\gamma^{t\rightarrow{tH}}$ that is the probability that a top quark emit one Higgs with momentum q.  Using equation (30) and refs[19,22,23] we calculate  the amount of this cross section :

\begin{eqnarray}
\gamma^{t\rightarrow{tH}}=\frac{2\alpha_s\alpha^2M_H^3}{6{\pi}{\nu}^2q^2}\left[-4C_F\left(\frac{M_t^2z\left(1-z\right)}{M_t^2\left(1-z\right)^2+M_H^2}\right)+2C_F\left(\frac{1+z^2}{1-z}\right)+2C_F\delta\left(1-z\right)\right]
\end{eqnarray}

\begin{eqnarray}
&&\sigma^{pp\rightarrow{t\overline{t}}} = 3\left(\frac{1}{3}\right)\left(\frac{1}{3}\right)\sum_{a}\int^{1}_{\tau'}dx_a\int^{1}_{\frac{\tau'}{x_a}}dx_{\overline{a}}f_{a/A}\left(x_a,\mu^2\right)\times f_{\overline{a}/B}\left(x_{\overline{a}},\mu^2\right) \sigma^{a\overline{a}\rightarrow{t\overline{t}}} \left(s x_{a} x_{\overline{a}} \right)
\delta \left(x_ax_{\overline{a}}-\frac{{q'_0}^2}{s}\right)\nonumber\\
&&~~~~~~~~~ =  N_{color}\frac{16{\pi}\alpha^2}{3{q'_0}^2}\left(\frac{s}{{q'_0}^2}\right)^{1.2}\left[1+3\frac{{q'_0}^2}{s}ln\left(\frac{s}{{q'_0}^2}\right)\right]\nonumber\\
&& \sigma^{ab\rightarrow{tt}}  = N_{color}\frac{4{\pi}\alpha^2}{3{q'_0}^2}e_t^2=N_{color}\frac{4{\pi}\alpha^2}{27{q'_0}^2}~~~~~~~~ for~~e_t=\frac{2}{3}\nonumber\\
&& \tau'=\frac{{q'_0}^2}{s}
\end{eqnarray}

\begin{eqnarray}
&& \frac{{d\sigma_1^{PQCD}}_{pp\rightarrow{H}}}{dq^2dz}=\sigma_{pp\rightarrow{t\overline{t}}}\gamma^{t\rightarrow{tH}}\nonumber\\
&& = N_{color}\frac{32{\pi}\alpha_{H}\alpha^2}{9{q'_0}^2}\frac{M_H^3}{\nu^2q^2}\left(\frac{s}{{q'_0}^2}\right)^{1.2}\left[1+3\frac{{q'_0}^2}{s}ln\left(\frac{s}{{q'_0}^2}\right)\right]\nonumber\\
&& \times\left[-4C_F\left(\frac{M_t^2z\left(1-z\right)}{M_t^2\left(1-z\right)^2+M_H^2}\right)+2C_F\left(\frac{1+z^2}{1-z}\right)-2C_F\delta\left(1-z\right)\right]
\end{eqnarray}

\begin{eqnarray}
&& {\sigma_{1}^{PQCD}}_{pp\rightarrow{H}}\nonumber\\ 
&& = \int_{M^2_{H}}^{Q_0^2}dq^2\int_{\frac{M_H^2}{q^2}}^{1-\frac{M_H^2}{q^2}}dz\frac{d\sigma_{pp\rightarrow{H}}}{dq^2dz}\nonumber\\
&& = \int_{M^2_{H}}^{Q_0^2}dq^2\int_{\frac{M_H^2}{q^2}}^{1-\frac{M_H^2}{q^2}}dz\sigma_{pp\rightarrow{t\overline{t}}}\gamma^{t\rightarrow{tH}}\nonumber\\
&& = \int_{M^2_{H}}^{Q_0^2}dq^2\int_{\frac{M_H^2}{q^2}}^{1-\frac{M_H^2}{q^2}}dzN_{color}\frac{32{\pi}\alpha_{H}\alpha^2}{9{q'_{0}}^2}\frac{M_{H}^3}{\nu^2q^2}\left(\frac{s}{{q'_0}^2}\right)^{1.2}\left[1+3\frac{{q'_0}^2}{s}ln\left(\frac{s}{{q'_0}^2}\right)\right]\nonumber\\
&& \times\left[-4C_F\left(\frac{M_t^2z\left(1-z\right)}{M_t^2\left(1-z\right)^2+M_H^2}\right)+2C_F\left(\frac{1+z^2}{1-z}\right)+2C_F\delta\left(1-z\right)\right]\nonumber\\
&& \approx \frac{32M_H^3\alpha_{H}\alpha^2}{9\nu^2{q'_0}^2}N_{color}\left(\frac{s}{4M_t^2}\right)^{1.2}\left[1+12\frac{M_t^2}{s}ln\left(\frac{s}{4M_t^2}\right)\right]\nonumber\\
&& \times\left[-2C_Fln\left(1-\frac{M_H^2}{4M_t^2}\right)+C_Fln\left(\frac{4M_t^2}{M_H^2}\right)-2C_F\frac{M_H^2}{M_t^2}arctan\left(\frac{M_t^2}{M_H^2}\right)\right]
\end{eqnarray}

where $M_t$ is the heavy quark mass ,$\sqrt{{q'_0}^2}=\sqrt{4M_t^2}$ is the invariant mass for the production of top quark-anti top quark pairs,  $\alpha_{H}$is the Higgs coupling,$\nu^2=\left(\sqrt{2}G_F\right)^{-1}\approx246GeV$, $\sqrt{s}$ is the centre of mass energy and $\alpha=\frac{e^2}{4{\pi}}$ . 
   Probably two protons annihilate and a pair of bottom quark anti bottom quark is produced. Then these particles interact with each other and create Higgs Boson. The cross section for this interaction is:

\begin{eqnarray}
&& {\sigma_{2}^{PQCD}}_{pp\rightarrow{H}}\nonumber\\
&& = \int\frac{d^4q_1}{\left(2{\pi}\right)^4}\int\frac{d^4q_2}{\left(2{\pi}\right)^4}2{\pi}{\delta}^4\left(\overrightarrow{k_1}+\overrightarrow{k_2}-\overrightarrow{q_1}-\overrightarrow{q_2}\right)L^{pp\rightarrow{b\overline{b}}}\left(k_1,k_2\right)p_q\left(q_1\right)p_{\overline{q}}\left(q_2\right)L^{b\overline{b}\rightarrow{H}}\left(q_1,q_2\right)\nonumber\\
&& \approx N_{color}\frac{2{\pi}\alpha^2}{9}\frac{\alpha_s}{\nu^2M_H^2}\left(\frac{s}{4M_b^2}\right)^{1.2}\left[1+12\frac{M_b^2}{s}ln\left(\frac{s}{4M_b^2}\right)\right]\left[-4C_Fln\left(1-\frac{M_b^2}{M_H^2}\right)+2C_Fln\left(\frac{M_b^2}{M_H^2}\right)+2C_F\right]\nonumber\\
&& \times\left[-ln\left(\frac{M_H-M_b}{M_H+M_b}\right)+\frac{M_H^2}{M_b^2}\right]
\end{eqnarray}

where we define the bottom quark part ,bottom cross section, Higgs part and bottom quark propagator for this integral as:

\begin{eqnarray}
&& L^{pp\rightarrow{b\overline{b}}}\left(k_1,k_2\right)\nonumber\\
&& = 3\left(\frac{1}{3}\right)\left(\frac{1}{3}\right)\sum_{a}\int_{\tau'}^{1}dx_a\int_{\frac{\tau}'{x_a}}^{1}dx_{\overline{a}}f_{a/A}\left(x_a,\mu^2\right)
\times f_{\overline{a}/B} \left(x_{\overline{a}},\mu^2 \right) 
\sigma^{a\overline{a}\rightarrow{b\overline{b}}} \left(s x_a x_{\overline{a}}\right)\delta \left(x_{a} x_{\overline{a}}-\frac{\left(k_1+ k_2\right)^2}{s}\right)\nonumber\\
&& = N_{color}\frac{4{\pi}\alpha^2}{3\left(k_{1} + k_{2} \right)^2} \left(\frac{s}{\left(k_{1}+ k_2 \right)^2}\right)^{1.2}\left[1+ 3\frac{\left(k_1+ k_2\right)^2}{s} \ln\left(\frac{s}{\left(k_1+ k_2\right)^{2}}\right)\right]
\end{eqnarray}

\begin{eqnarray}
\sigma^{ab\rightarrow{b\overline{b}}}=N_{color}\frac{4{\pi}\alpha^2}{3x_ax_{\overline{a}}s}e_{b}^2=N_{color}\frac{4{\pi}\alpha^2}{27x_ax_{\overline{a}}s}e_{b}^2 ~~~~~~~~~~~~ for~ e_b=-\frac{1}{3}
\end{eqnarray}

\begin{equation}
L^{b\overline{b}\rightarrow{H}}\left(q_1,q_2\right)=\frac{\left(\overrightarrow{q_1}+\overrightarrow{q_2}\right)^2\alpha_s}{24\nu^2M_H^2}\left[-4C_Fln\left(1-\frac{M_H^2}{\left(\overrightarrow{q_1}+\overrightarrow{q_2}\right)^2}\right)+2C_Fln\left(\frac{\left(\overrightarrow{q_1}+\overrightarrow{q_2}\right)^2}{4M_H^2}\right)+2C_F\right]
\end{equation}

\begin{equation}
P_b=\frac{\delta^{ab}}{q^{\mu}\gamma_{\mu}-m}
\end{equation}

    The perturbative cross sections for Higgs boson production increase with center of mass energy and decrease by increasing Higgs boson mass. Thus we expect that many Higgs bosons produce at TeV center of mass energies. The production cross section of Higgs boson resulting from quark-anti quark interactions depends on strong coupling constant. 
We conclude that the cross section for Higgs radiation from black hole is dominated by PQCD, since these particles produce via different channels near event horizon of mini black holes at LHC. The number of these channels is more than those in PQCD. When black holes produce at LHC, they evaporate to Massive particles like Higgs boson. In fact black holes at LHC act as a factory for Higgs production. Furthermore small black holes decay to bottom-anti bottom. These particles interact with each other and produce Higgs boson. Also many Higgs bosons are produced due to gluon-gluon interaction.


\section{The effect of Higgs boson production inside the TeV black holes on hadronic cross section}
In this section we analyze the effect of Higgs boson production inside the shell and outside the shell of mini black holes on hadronic cross section at LHC.If mini black holes In this section we analyze the effect of Higgs boson production inside the shell and are produced at LHC, many Higgs bosons produce and decay to QCD matter near them and this will result in a difference between the observed hadronic cross sections and the predicted cross sections. As we discussed in section IV,the number of channels for producing these particles is more than the corresponding number of channels in PQCD. Also many Higgs bosons produce directly inside TeV black holes and decay to quark-antiquark. The quarks or antiquarks may emit gluons and form three jet events.
Using equations(31,36,39,43,46,48,52,53) and the results in ref[1] we can calculate the hadronic cross section production:

\begin{eqnarray}
&& \sigma_3=\int{dq'^2}2{\pi}\delta \left(Q_0^2-q'^2\right)\int_{M_H^2}^{q'^2}dq^2 \int_{1-\frac{q^2}{q'^2}}^{\frac{q^2}{q'^2}}dz \nonumber \\
&& ~~~\times \bigg[\sigma^1_{pp \rightarrow{H},{outside~the~shell}} \left(q'\right)+ \sigma^2_{pp\rightarrow{H},{outside~the~shell}}\left( q' \right)+ \sigma^3_{pp \rightarrow{H},{outside~the~shell}} \left( q' \right) \nonumber\\
&& ~~~+ \sigma^{1}_{pp \rightarrow {H},{inside~the~shell}} \left(q'\right) 
+ \sigma^{2}_{pp \rightarrow {H}, {inside~the~shell}} \left(q'\right)+ \sigma^{3}_{pp \rightarrow{H}, {inside~the~shell}} \left(q'\right) \nonumber\\
&& ~~~+\sigma^1_{pp\rightarrow{H},outside~the~horizon}\left(q'\right)+\sigma^2_{pp\rightarrow{H},outside~the~horizon}\left(q'\right)+\sigma^3_{pp\rightarrow{H},outside~the~horizon}\left(q'\right)\nonumber\\
&& ~~~+ {\sigma_{1}^{PQCD}}_{pp \rightarrow{H}} \left( q' \right)+ {\sigma_{2}^{PQCD}}_{pp \rightarrow{H}} \left( q' \right) \bigg] \Gamma^{ H \rightarrow{q \overline{q}}} \left( q \right) \gamma^{q \rightarrow{qg}} \left( z \right)
\end{eqnarray}

where $\sqrt{Q_0^2}=\sqrt{s}$  is the center of mass energy, $\Gamma^{H\rightarrow{q\overline{q}}}$  is the rate for decay of Higgs boson to quark antiquark, [22],z is the fraction of  momentum quark or antiquark that transforms to gluon momentum, $q'$ is the sum of  particle energies that produce in proton- proton collisions and $\gamma6{q\rightarrow{qg}}$ [19] is the probability for gluon emission  from quark,
\begin{equation}
\gamma^{q\rightarrow{qg}}=\frac{\alpha_s}{3{\pi}}\left[-4C_F\left(\frac{M_q^2z\left(1-z\right)}{M_q^2\left(1-z\right)^2+q^2}\right)+2C_F\left(\frac{1+z^2}{1-z}\right)+2C_F\delta\left(1-z\right)\right]
\end{equation}

\begin{equation}
\Gamma^{H\rightarrow{q\overline{q}}}=\frac{2C_F}{q^2}\left[-2ln\left(1-\frac{M_H^2}{q^2}\right)+ln\left(\frac{M_q^2}{M_H^2}\right)+2\right]
\end{equation}

We should regard all probable energies for quark anti quark and gluon production. For this reason we integrate over the range of energy from $M_H$ to $\sqrt{s}$  .We can approximate the integral in equation(58)as:

\begin{equation}
\sigma_3=\alpha_sA+\left(\alpha_s\right)^2B+\left(\alpha_s\right)^3C
\end{equation}

where we define the coefficients A,B and C as:

\begin{eqnarray}
&& A\approx \left(f+1\right)\times\frac{C_F}{12{\pi}M_{BH}}\left[ln\left(\frac{M_{BH}^2}{M_{H}^2}\right)ln\left(\frac{M_{BH}^2}{M_H^2}-1\right)\right]\frac{1}{M_P^2}\left[\frac{M_{BH}}{M_P}\left(\frac{8\Gamma\left(\frac{7}{2}\right)}{6}\right)\right]^{\frac{2}{5}}\left(\frac{Q_0^2}{M_{BH}^2}\right)^{1.2} \nonumber\\
&&~~ \times \left[1+3\frac{M_{BH}^2}{Q_0^2}ln\left(\frac{Q_0^2}{M_{BH}^2}\right)\right] + f \times \left[8{\pi}\left(\frac{Q_0^2}{M_{BH}}\right)^{1.2}-\left(\frac{Q_0^2}{M_{BH}^3}\right)^{1.2}\right] \nonumber\\
&&~~ \times \left[1+ 3\frac{M_{BH}^2}{s} ln \left(Q_0^2 \right)+ 3\frac{M_{BH}^3}{s}\left(1- ln\left(M_{BH}^2\right) \right)\right] \frac{1}{M_P^2}\left[\frac{M_{BH}}{M_P}\left(\frac{4\Gamma\left(\frac{7}{2}\right)}{3}\right)\right]^{\frac{2}{5}}\nonumber\\
&&~~ \times \left[ \frac{M_{BH}^2}{M_{h}^2}\left( ln\left( \frac{M_{BH}^2}{M_{H}^2}\right)+ 1\right)+ \left( \frac{M_{BH}^2}{M_{H}^2}- 1\right) \left(ln \left( \frac{M_{BH}^2}{M_{H}^2}- 1\right)+ 1\right) \right] \nonumber\\
&&~~ \times \left[2+ ln\left(\frac{M_q^2}{M_H^2}\right)+ \frac{1}{2M_H^2} ln\left( 1- \frac{M_H^2}{Q_0^2} \right)^2- \frac{1}{2M_H^2} ln \left(1- \frac{M_H^2}{M_{BH}^2} \right)^2 \right] \nonumber\\
&&~~ \times \left[ ln\left( \frac{Q_0^2}{M_{BH}^2} \right)-ln\left( \frac{M_q^2}{Q_0^2} \right)+o\left( higher \right) \right]
\end{eqnarray}

\begin{eqnarray}
&& B \approx  \left( f+ 1 \right) \frac{M_b^2}{24{\nu}^2M_H^2} N_{color}\left( \frac{Q_0^2}{M_{BH}^2} \right)^{1.2} \left[ 1+ 3\frac{M_{BH}^2}{Q_0^2} ln\left( \frac{Q_0^2}{M_{BH}^2} \right) \right] \left[-4C_F ln\left(1- \frac{M_H^2}{M_{BH}^2} \right)+2C_F ln\left( \frac{1}{4} \right)+ 2C_F \right]\nonumber\\
&&~~ \times  \frac{1}{M_P^2} \left[ \frac{M_{BH}}{M_P} \left( \frac{4\Gamma \left( \frac{7}{2} \right)}{3} \right) \right]^{\frac{2}{5}} \frac{e^{-8{\pi}M_{BH}^2}}{1+e^{-8{\pi}M_{BH}^2}} \left[-ln\left( \frac{M_H-M_b}{M_H+M_b} \right)+ \frac{M_H^2}{M_b^2} \right]+ \frac{8M_H^3\alpha_{s}}{9M_t^2} N_{color} \left( \frac{Q_0^2}{4M_t^2}\right)^{1.2} \nonumber\\
&&~~ \times \left[ 1+ 12\frac{M_t^2}{Q_0^2}ln\left( \frac{Q_0^2}{4M_t^2} \right) \right] \left[ -2C_Fln \left( 1- \frac{M_H^2}{4M_t^2} \right)+ C_Fln \left( \frac{4M_t^2}{M_H^2} \right) -2C_F \frac{M_H^2}{M_t^2}arctan \left( \frac{M_t^2}{M_H^2} \right) \right] \nonumber\\
&&~~ + N_{color} \frac{2{\pi} \alpha^{2}}{9} \frac{1}{{\nu}^{2} M_H^{2}} \left( \frac{Q_0^2}{4M_b^2} \right)^{1.2} \left[ 1+ 12\frac{M_b^2}{Q_0^2}ln\left( \frac{Q_0^2}{4M_b^2} \right) \right] \left[ -4C_Fln \left(1-\frac{M_b^2}{M_H^2} \right)+ 2C_Fln\left( \frac{M_b^2}{M_H^2} \right)+ 2C_F\right] \nonumber\\
&&~~ \times \left[ -ln\left( \frac{M_H-M_b}{M_H+M_b} \right) + \frac{M_H^2}{M_b^2} \right]+ \left[ f \times {N_{color}} \delta_{cd} \left( 8{\pi} \left( \frac{Q_0^2}{M_{BH}} \right)^{1.2}+ \left( \frac{Q_0^2}{M_{BH}^3} \right)^{1.2} \right)\right] \left[ 1+ 3\frac{M_{BH}^2}{Q_0^2} ln\left( Q_0^2 \right)\right] \nonumber\\
&&~~ + \left[ 3 \frac{M_{BH}^3}{s} \left( 1-ln \left( M_{BH}^2 \right) \right) \right] \frac{1}{M_P^2} \left[ \frac{M_{BH}}{M_P} \left( \frac{4\Gamma \left( \frac{7}{2} \right)}{3} \right) \right]^{\frac{2}{5}} \left[ -ln\left( \frac{M_H-M_b}{M_H+M_b} \right)+ \frac{M_H^2}{M_b^2} \right]  \nonumber\\ 
&&~~ \times \frac{M_b^2 \alpha_{s}}{24{\nu}^2M_H^2} \left[-4C_F M_{BH}^2\left(1-ln \left(1-\frac{M_H^2}{M_{BH}^2} \right) \right) \right]++2C_Fln\left(\frac{1}{4}\right) \nonumber\\
&&~~ + 2C_F \left[ 2+ ln\left( \frac{M_q^2}{M_H^2} \right)+ \frac{1}{2M_H^2} ln\left( 1-\frac{M_H^2}{Q_0^2} \right)^2- \frac{1}{2M_H^2}ln\left( 1- \frac{M_H^2}{M_{BH}^2} \right)^2 \right] \nonumber\\
&&~~ \times \left[ -ln\left( \frac{M_{BH}^2}{Q_0^2} \right)-ln\left( \frac{M_q^2}{Q_0^2} \right)+ 4ln\left( \frac{Q_0^2-M_H^2}{M_{BH}^2-M_H^2} \right)+ \frac{2}{3}ln^{3} \left( \frac{M_{BH}^2}{Q_0^2} \right)+ 0\left( higher \right) \right]
\end{eqnarray}

\begin{eqnarray}
&& C \approx \left(f+1\right)\times{C_f}\left(\frac{Q_0^2}{M_{BH}^2}\right)^{1.2}\left[1+3\frac{M_{BH}^2}{Q_0^2}ln\left(\frac{Q_0^2}{M_{BH}^2}\right)\right]\frac{{\pi}}{288\sqrt{2}} \nonumber\\
&&~~ \times \left[\frac{6M_t^2}{M_H^2}\left(1+\left(1-\frac{4M_t^2}{M_H^2}\right)arcsin^2\left(\sqrt{\frac{M_H^2}{4M_t^2}}\right) \right) \right]^2 \times \frac{1}{M_P^2}\left[\frac{M_{BH}}{M_P}\left(\frac{4\Gamma\left(\frac{7}{2}\right)}{3}\right)\right]^{\frac{2}{5}}\frac{e^{-8{\pi}M_{BH}^2}}{1+e^{-8{\pi}M_{BH}^2}} \nonumber\\
&&~~ \times \left[\frac{1}{M_t^2}-\frac{1}{M_H^2}+ln\left(\frac{M_t}{M_H}\right)\right]+\left[f \times C_F \delta_{mn}\left(8{\pi}\left(\frac{Q_0^2}{M_{BH}}\right)^{1.2}-\left(\frac{Q_0^2}{M_{BH}^3}\right)^{1.2}\right)\right] \nonumber\\ 
&&~~ \times \left[ 1+ 3\frac{M_{BH}^2}{Q_0^2} ln\left( Q_0^2 \right)+ 3\frac{M_{BH}^3}{Q_0^2} \left( 1- ln\left( M_{BH}^2 \right) \right) \right] \frac{1}{M_P^2} \left[ \frac{M_{BH}}{M_P} \left( \frac{4\Gamma \left( \frac{7}{2} \right)}{3} \right) \right]^{\frac{2}{5}} \nonumber\\ 
&&~~ \times \left[ \frac{1}{M_t^2}- \frac{1}{M_H^2}+ ln\left( \frac{M_t}{M_H} \right) \right] \left( \frac{\alpha_{s}}{\pi} \right)^2 \frac{\pi}{288\sqrt{2}} \left[ \frac{6M_t^2}{M_H^2} \left( 1+\left( 1- \frac{4M_t^2}{M_H^2} \right) arcsin^2 \left( \sqrt{\frac{M_H^2}{4M_t^2}} \right) \right) \right]^2 \nonumber\\
&&~~ \times \left[ 2+ ln\left( \frac{M_q^2}{M_H^2} \right)+ \frac{1}{2M_H^2} ln\left( 1- \frac{M_H^2}{Q_0^2} \right)^2 -\frac{1}{2M_H^2} ln\left( 1-\frac{M_H^2}{M_{BH}^2} \right)^2 \right] \nonumber\\
&&~~ \times \left[ -ln\left( \frac{M_{BH}^2}{Q_0^2} \right)- ln\left( \frac{M_q^2}{Q_0^2} \right)+ 4ln\left( \frac{Q_0^2-M_H^2}{M_{BH}^2-M_H^2} \right)- \frac{2}{3} ln^3\left( \frac{M_{BH}^2}{Q_0^2} \right) \right] \nonumber\\
&&~~ + \left[ \frac{Q_0^2}{M_q^2}arctan\left( \frac{M_q^2}{Q_0^2} \right)- \frac{Q_0^2-M_q^2}{M_{BH}^2-M_q^2} \left(ln \left( \frac{Q_0^2-M_q^2}{M_{BH}^2-M_q^2} \right) +1 \right) + 0 \left( higher \right) \right]
\end{eqnarray}

\begin{figure}
\includegraphics[scale=1.3]{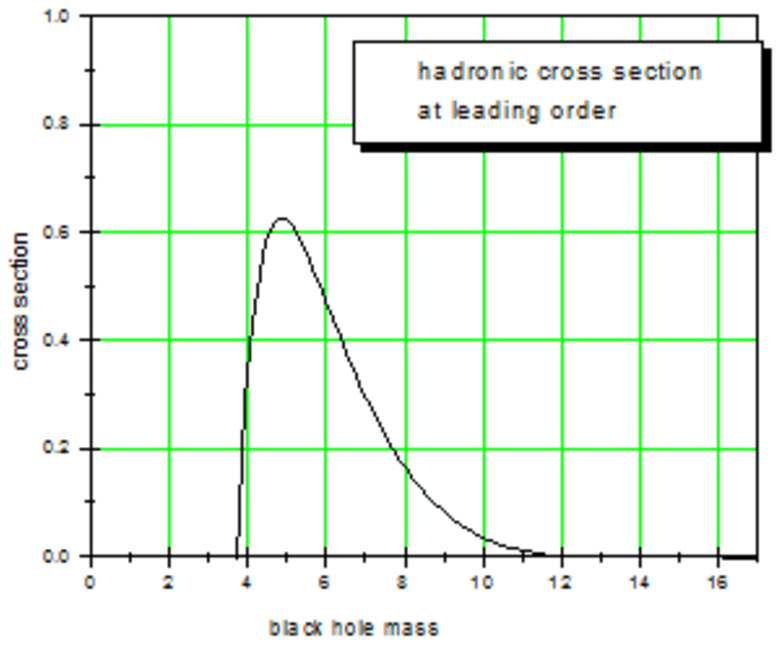}
\caption{The hadronic cross sections a function of black hole mass at Leading Order(A coefficient).}
\label{Fig. 8}
\end{figure}

\begin{figure}
\includegraphics[scale=1.3]{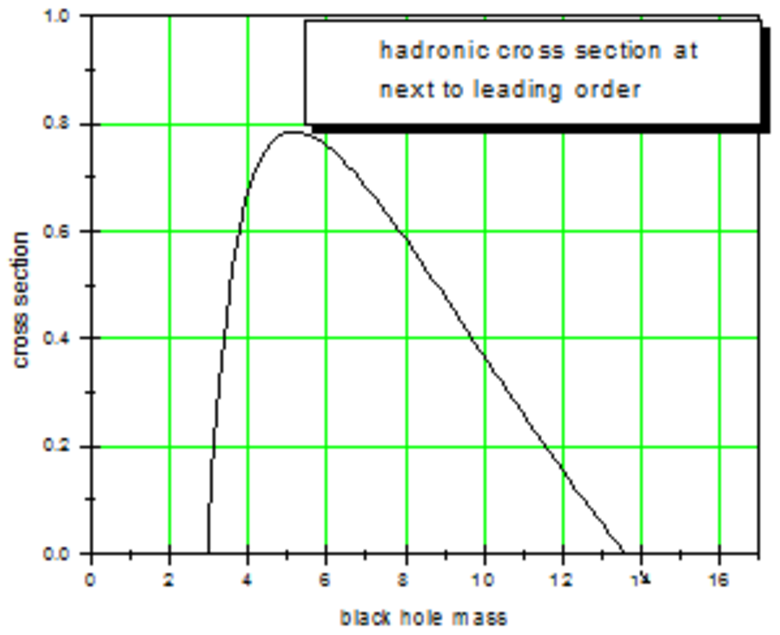}
\caption{The hadronic cross sections a function of black hole mass at Next to Leading Order(B coefficient).}
\label{Fig. 9}
\end{figure}

\begin{figure}
\includegraphics[scale=1.3]{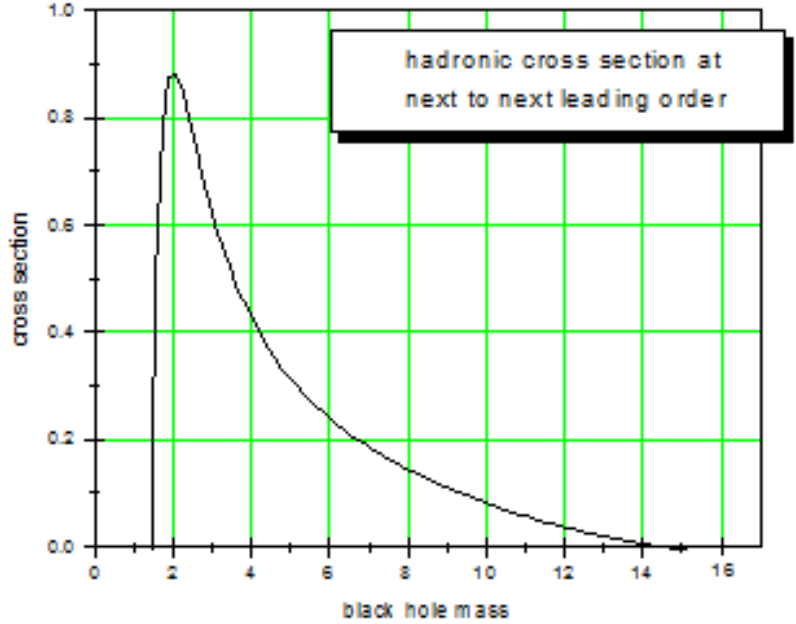}
\caption{The hadronic cross sections a function of black hole mass at Next to Next to Leading Order(C coefficient).}
\label{Fig. 10}
\end{figure}

\pagebreak
In figures(8-10) we show hadronic cross sections at Leading Order LO, Next to Leading Order, NLO, and Next to Next Leading order, NNLO respectively. In these Figures we choose $M_H=130GeV ,M_P=2TeV,M_t=175GeV, M_b=5GeV$  and $\sqrt{s}=14TeV$  for Higgs boson mass, Planck mass,top quark mass ,bottom quark mass and center of mass energy, respectively.In each figure,the effect of Higgs boson on hadronic cross section decreases at LHC as the mass of black hole increases.We observe that as the order of perturbation theory increases, the effect of Higgs boson radiation from mini black holes on hadronic cross section becomes systematically more effective, because at higher orders there exists more channels for Higgs production and their decay of Higgs into massive quark-anti quarks in our calculation. On the other hand at lower mass, the black hole will not able to emit heavy particles and the Higgs boson cross section decreases. This leads to a decrease in quark-antiquark production via Higgs boson decay to QCD matter and consequently to a decrease in hadronic cross section. At smaller mass$(M_{BH}<2TeV )$,the NNLO contribution is large while the cross sections at NLO and at LO are rising at $M_{BH}<3 TeV and M_{BH}>4 TeV$ respectively and exhibit a turn-over at moderate values of  black hole mass. The peak moves from about $5TeV$ (LO) to $2.5TeV$ (NNLO). These results are different from ref[1] due to effects of Higgs boson production and decay inside the event Horizon.It’s conclude that the processes of hadronization inside the event horizon of mini black holes is affected the hadronic cross section outside the event horizon and can be observed at LHC.

\section{Summary and Outlook}

In this paper first we consider different channels for Higgs production from inside and outside the shell of TeV black holes at proton proton annihilation .Then we compare the cross section for these interactions with PQCD.Finally we predict the effect of Higgs boson production inside the mini black holes on three-jet rate at LHC by summing over all cross sections for Higgs boson radiation from black hole and also from Higgs decay in PQCD.

\section{References:}
$[1]$ A.sepehri, M.E.Zomorrodian, A.Moradi Marjeneh, P.Eslami, S.Shoorvazi, Can. J. Phys.90: 25-37 (2012)\\
$[2]$ Zomorrodian, M.E., Sepehri, A., Marjaneh, A.M. , Can. J. Phys. 88(11):841-849(2010)\\
$[3]$ Landsberg, Greg L, PhysRevLett.88.181801\\
$[4]$ Andrew Chamblin, Fred Cooper, Gouranga C.Nayak, Phys.Lett.B672:147-151(2009)\\
$[5]$ G. C. Nayak and J. Smith, Phys. Rev. D 74, 014007 (2006)\\
$[6]$ A. Chamblin, F. Cooper and G. C. Nayak, Phys. Rev. D 70, 075018 (2004).\\
$[7]$ W. Beenakker, R. H¨opker, M. Spira and P.M. Zerwas, Nucl. Phys. B492, 51 (1997)\\
$[8]$ L. Anchordoqui and H. Goldberg, Phys. Rev. D 67, 064010 (2003). R. Emparan, G. T. Horowitz and R. C. Myers, Phys. Rev. Lett. 85, 499 (2000).\\
$[9]$ P. Kanti and J. March-Russell, Phys. Rev. D 67, 104109 (2003). A. Chamblin, F. Cooper and G. C. Nayak, Phys. Rev. D 69, 065010 (2004).\\
$[10]$ Doyeol Ahn, Phys. Rev. D74: 084010, (2006). W. G. Unruh, Phys. Rev. D 14, 870 (1976).\\
$[11]$ D. Ahn, Y.H. Moon, R. B. Mann, I. Fuentes-Schuller, JHEP0806:062, (2008).\\
$[12]$ C. Itzykson and F-B. Zuber, Quantum Field Theory, McGraw-Hill, New York (1980).\\
$[13]$ Jun-Chen Su, J. Phys. G27 (2001) 1493-1500\\
$[14]$ N. D. Birrell and P. C. W. Davies, Quantum fields in curved space (Cambridge University Press, New York, 1982). \\
$[15]$ Xavier Calmet, Wei Gong, Stephen D. H. Hsu, Phys.Lett.B668:20-23,2008.\\
$[16]$ Douglas M. Gingrich, J. Phys. G: Nucl. Part. Phys. 37 (2010) 105008.\\
$[17]$ T. Ghaffary, Eur. Phys. j. A 53:20 (2017)\\
$[18]$ T. Banks, W. Fischler, arXiv:hep-th/9906038. S. B. Giddings and S. Thomas, Phys. Rev. D 65, 056010 (2002)\\
$[19]$ M. Spira, A. Djouadi, D. Graudenz and P. M. Zerwas, Nucl. Phys. B 453 (1995) 17\\
$[20]$ Pushpalatha C. Bhat, Russell Gilmartin, Harrison B. Prosper, Phys.Rev. D62 (2000) 074022\\
$[21]$ S.Lloyd.Phys.Rev.Lett.96, 061302(2006).doi:10.1103 /PhysRevLett.96.061302.PMID:16605980.\\
$[22]$ Nikolaos Kidonakis, Phys.Rev.D77:053008,(2008). Mathias Butenschoen, Frank Fugel, Bernd A. Kniehl, Phys.Rev.Lett.98:071602,(2007).\\
$[23]$ German Rodrigo, Frank Krauss Eur.Phys.J. C33 (2004) .S457-S459, Frank Krauss, German
Rodrigo, Phys.Lett. B576 (2003) 135-142\\
\end{document}